\documentclass{emulateapj}
\usepackage{apjfonts}
\usepackage[breaklinks,colorlinks=true,linkcolor=blue,anchorcolor=blue,citecolor=blue,urlcolor=blue,draft=false]{hyperref}

\usepackage{graphics,graphicx,rotating}
\usepackage{xspace}
\usepackage{amssymb}
\usepackage{amsmath}
\usepackage{epstopdf}
\usepackage{subfigure}
\usepackage{amsthm,amscd}
\usepackage{color}
\usepackage{soul}
 \usepackage{textcomp}
\usepackage{gensymb}

\shorttitle{Test for relativistic kinetic theories based on the Sunyaev-Zel'dovich effect}
\shortauthors{}

\newcommand{\simless} 
     {\ensuremath{\lower 3pt\hbox{$\rlap{\raise5pt\hbox{$\char'074$}}\mathchar"7218$}}}
\newcommand{\simgreat}
     {\ensuremath{\lower 3pt\hbox{$\rlap{\raise5pt\hbox{$\char'076$}}\mathchar"7218$}}}

\newcommand{\simgt}{\lower.5ex\hbox{$\; \buildrel > \over \sim \;$}}
\newcommand{\simlt}{\lower.5ex\hbox{$\; \buildrel < \over \sim \;$}}

\newcommand{\KMSEC}{{$\rm km\;s^{-1}$}\xspace}

\newcommand*{\ltsim}{\ {\raise-.75ex\hbox{$\buildrel<\over\sim$}}\ }
\newcommand*{\gtsim}{\ {\raise-.75ex\hbox{$\buildrel>\over\simsaus$}}\ }
\newcommand*{\proptosim}{\ {\raise-.75ex\hbox{$\buildrel\propto\over\sim$}}\ }

\newcommand*{\NBODYHYD}{{$N$-body/hydro\-dynamical}\xspace}

\newcommand*{\vmath}{\textsl{v}}

\newcommand*{\CHANDRA}{\emph{Chandra}\xspace}

\newcommand*{\FLASH}{\textsc{flash}\xspace}

\newcommand{\FIGURES}{./}

\newcommand*{\JUTTNER}{J\"{u}ttner\xspace}
\newcommand*{\PLANCK}{{\it Planck}\xspace}
\newcommand*{\HERSCH}{{\it Herschel}\xspace}

\begin{document}

\title{Empirical test for relativistic kinetic theories based on \\ the Sunyaev-Zel'dovich effect}

\author{
S. M. Molnar\altaffilmark{1} and J. Godfrey\altaffilmark{2}
}

\altaffiltext{1}{Institute of Astronomy and Astrophysics, Academia Sinica, 
                 No. 1, Section 4, Roosevelt Road, Taipei 10617, Taiwan, R.O.C.}

\altaffiltext{2}{Virginia Tech, System Performance Laboratory, 7054 Haycock Road, Falls Church, VA 22043-2311}


\begin{abstract}
We propose a new method to determine the electron velocity (EV) distribution function
in the intracluster gas (ICG) in clusters of galaxies based on the frequency dependence 
of the Sunyaev-Zel'dovich (SZ) effect. 
It is generally accepted that the relativistic equilibrium EV distribution
is the one suggested by \JUTTNER. 
However, there is an ongoing debate on the foundation of relativistic kinetic theory, 
and other distributions have also been proposed. 
The mildly relativistic intracluster gas (ICG) provides a unique laboratory to test 
relativistic kinetic theories.
We carried out Monte Carlo simulations to generate SZ signal from a 
single-temperature gas assuming the \JUTTNER EV distribution assuming a few per cent errors.
We fitted SZ models based on non-relativistic Maxwellian, 
and its two relativistic generalizations, the \JUTTNER and modified \JUTTNER distributions. 
We found that a 1\% error in the SZ signal is sufficient to distinguish between these 
distributions with high significance based on their different best-fit temperatures.
However, in any LOS in a cluster, the ICG contains a range of temperatures.
Using our \NBODYHYD simulation of a merging galaxy cluster
and assuming a 1\% error in the SZ measurements in a LOS through a bow shock, 
we find that it is possible to distinguish between \JUTTNER and modified \JUTTNER 
distributions with high significance. 
Our results suggest that deriving ICG temperatures from fitting to SZ data
assuming different EV distribution functions and comparing them 
to the temperature in the same cluster obtained using other observations
would enable us to distinguish between the different distributions.
\end{abstract}

\section{Introduction}
\label{S:Intro}

The generalization of non-relativistic kinetic theory to relativistic velocities is still 
a subject of debate.
Presently, there does not exist a theory of relativistic statistical mechanics that can 
account for the approach to equilibrium of an ensemble of particles, i.e., a gas,
or the relativistic generalization of the many-body problem (for a review see \citealt{Hakim2011}).
The essential difficulty is that, unless all particles in the ensemble originate from
the same point in space, they are initially space-like separated. 
Therefore neither a time-ordering of the particle states, 
nor their initial conditions can be established.
These difficulties impede developing a microscopic theory for how a relativistic gas can achieve 
thermal equilibrium. 
Also, treating many-particle interactions within the framework of special relativity
in kinetic theories is problematic.
Even if we assume that the relativistic gas is in equilibrium, 
the statistical mechanical treatment of the gas fails because interactions cannot be 
included in the relativistic Hamiltonian in a consistent manner, 
since any interaction term in the Hamiltonian would break the Poincar\'e symmetry
(non-interaction theorems, e. g., \citealt{Leutwyler1965,MarmoET1984}).
This is not a problem in nonrelativistic statistical mechanics, 
because it allows the treatment of interactions between particles 
based on instantaneous interaction potentials, which are 
additive in the corresponding Hamiltonian 
straightforwardly leading to equilibrium velocity distributions.

In 1911, Ferencz \JUTTNER derived a relativistic generalization of the non-relativistic
(Maxwell) velocity distribution, usually referred to as the \JUTTNER, or Maxwell-\JUTTNER
velocity distribution \citep{Juttner1911}. 
For a general introduction to relativistic statistical mechanics, see \cite{Synge1957}.
However, problems with the establishment of a self-consistent relativistic kinetic theory 
led to the question whether the \JUTTNER distribution is the correct 
relativistic equilibrium velocity distribution, and other, 
modified \JUTTNER distributions were suggested (e.g., 
\citealt{DunkHang2007PhyA374,Kaniadakis2006,Lehmann2006JMP47,Schieve2005,HorwitzET1989PhyA161}).
Relativistic molecular dynamics Monte Carlo 
simulations in one, two, and three dimensions seem to 
support the \JUTTNER distribution for equilibrium
\citep{CuberoET2007,MontakhabET2009,PeanoET2009,Dunkel2009NatPh}.
These different approaches to relativistic kinetic theory make predictions, in principle testable,
and, as we show in this paper, feasible tests are possible.
As of today, there is no experimental verification of any of these proposed velocity distributions.

The correct relativistic equilibrium velocity distribution is essential in a variety of
applications, including high energy physics, astrophysics, and cosmology 
(e.g., \citealt{{DunkelET2007NJPh9},HeesET2006PhRvC,Bernstein2004,NozawaET1998ApJ507,
Rephaeli1995ApJ445}).
Therefore, it is of fundamental importance to find experimental or observational methods 
to derive particle velocity distribution in mildly relativistic gases, 
which would make it possible to distinguish between the proposed distributions.
In this paper we focus on experimental tests of the relativistic Maxwell-Boltzmann 
distribution as derived by \cite{Juttner1911}, \cite{DunkHang2007PhyA374} and 
\cite{HorwitzET1981}.

The high temperature low density gas of the intracluster gas (ICG) in clusters of galaxies
is mildly relativistic, the temperature is not high enough for particle pair creation and 
annihilation to be important. 
Thus the ICG provides a unique laboratory to test the proposed velocity distribution functions.

Inverse Compton scatterings of low energy cosmic microwave photons
off hot electrons in the ICG, the thermal Sunyaev-Zel'dovich (SZ) effect \citep{SZ1980ARAA18}, 
provides a possibility to study electron velocity (EV) distributions observationally.
Calculations of the relativistic SZ effect assume that the EV distribution is in the
form of \JUTTNER distribution (for reviews see, e.g., \citealt{Birkinshaw1999,Rephaeli1995ApJ445}.
However, as of today, there is no observational constraints on the EV distribution
function in the ICG.
\cite{ProkhorovET2011AA529} suggested that it may be possible to constrain the 
EV distribution in the ICG based on the frequency dependence of the SZ effect,
and estimated the accuracy needed in the SZ observations.
They considered Maxwellian and its relativistic generalization, the \JUTTNER 
EV distribution functions.
Their method compares the shape of the frequency dependence 
of the SZ effect based on different EV distribution functions.
Prokhorov et al.'s results suggest that, applying their method, 
SZ measurements with 0.1\% accuracy would be necessary to distinguish 
between Maxwellian and \JUTTNER velocity distributions.

In this paper we propose a new method to constrain 
EV distribution functions based on the thermal SZ effect and 
carry out Monte Carlo simulations to estimate the accuracy 
necessary to distinguish between the different EV distributions.
The structure of this paper is as follows. 
In Section~\ref{S:RELKIN} we give a short discussion about
the different approaches to relativistic kinetic theory.
We introduce the non-relativistic and relativistic thermal SZ effects
in Sections~\ref{S:NORELSZ} and \ref{S:RELSZ}.
In Section~\ref{S:RELSZ} we also discuss the differences between 
these approaches and highlight some less known subtleties
hidden in their assumptions.
We quantify the differences between non-relativistic 
and relativistic SZ effect at the wide frequency range 
adopted by the detectors on the \PLANCK satellite as well.
In Section~\ref{S:SZBULLET} we fit SZ models 
to SZ observations of the Bullet cluster based on 
EV distributions of the form of non-relativistic (Naxwellian), 
relativistic \JUTTNER and modified \JUTTNER.
We estimate the accuracy we need in SZ measurements to
constrain EV distribution functions with high significance
using mock SZ observations in Section~\ref{S:SZMC}.
We show our results for fits to mock SZ observations of a 
single-temperature gas in Section~\ref{SS:FITPLASMA}.
We fit SZ models to mock observations in a LOS through a 
bow shock extracted from an \NBODYHYD simulation based 
on EV distributions of the form of relativistic \JUTTNER and 
modified \JUTTNER in Section~\ref{SS:FITSHOCK}.
In Section~\ref{S:Discussion} we discuss our results and 
the feasibility of our new method to constrain EV distribution 
functions based on the SZ effect.
Section~\ref{S:Conclusion} contains our conclusion.

\section{Different approaches to relativistic kinetic theory}
\label{S:RELKIN}

As we mentioned in the Introduction, a relativistic generalization of classical 
kinetic theory cannot approach equilibrium, yet, we observe in the universe relativistic 
gas that we believe should be treated as in equilibrium.
The difficulties of establishing a self-consistent relativistic kinetic theory can be avoided 
if we simply assume that a system of relativistic particles, constituting a gas, is in equilibrium, 
and derive the equilibrium velocity distribution function using macroscopic methods, 
the principle of maximum entropy, as it was done by \JUTTNER \citep{Juttner1911}.
The maximum entropy principle (MEP) is particularly apt in this context precisely because we cannot 
give a microscopic account for how equilibrium is achieved. 
In a mildly relativistic gas the particle interactions are not so energetic as to significantly 
involve consideration of pair-creation, there are only two constraints on the entropy:
1) particle number conservation and 
2) energy conservation. 
Having formed a relativistically invariant Lagrangian, and 
carefully choosing an invariant measure for the phase space, 
the \JUTTNER distribution is obtained \citep{Synge1957}.

The standard classical macroscopic theory assumes
the Gibbs entropy functional, 

\begin{equation} \label{E:GIBBSEPY}
  S_{NR} [f] = - \int d^3 {\bf v} \;f({\bf v}) \log \{ f({\bf v}) \}
,
\end{equation}
where ${\bf v}$ is the 3D velocity.
The  velocity distribution function is derived by maximizing 
$S_{NR}$ under the constraints that $ f \ge 0$, 
$\int d^3 {\bf v} f({\bf v}) = 1$, and the energy conservation, 
$\langle E \rangle = \int d^3 {\bf v} \, f({\bf v}) \, E({\bf v})$, 
where the non-relativistic energy,  $E( {\bf v} ) = m\,  {\bf v}^2/2$.
Note, however, this method works only in Cartesian
coordinates, it is not coordinate invariant.
Using the relative entropy, the MEP can be cast in a coordinate invariant form
\citep{DunkelET2007NJPh9}.
Assuming a reference measure of $\rho  ({\bf p})$,
the relative entropy of $\Phi ({\bf p})$ with respect to the reference measure
can be written as 

\begin{equation} \label{E:RELENT}
  S [\Phi | \rho] = - \int d^3 p  \;\Phi({\bf p}) \log \{ \Phi({\bf p})/\rho({\bf p}) \}
,
\end{equation}
under the constraints that $\Phi \ge 0$,  $\rho \ge 0$, 
$\int d^3 p \; \Phi({\bf p}) = 1$, and 
$\langle E \rangle = \int d^3 p \; \Phi({\bf p}) \; E( {\bf p} )$, 
where the relativistic energy, $E( {\bf p} )^2 = m^2 c^4 + {\bf p}^2$. 
\cite{DunkelET2007NJPh9} showed that adopting 
a reference measure of $\rho_0 = 1/(m c)^3$, and 
maximizing the relative entropy, $S[\Phi | \rho_0]$ (Equation~\ref{E:RELENT}), 
one obtains the \JUTTNER distribution, 

\begin{equation} \label{E:RELJUTMJ}
  \Phi_\eta  ( {\bf p} )  = \frac{ \exp \{ - \beta E({\bf p}) \} }{Z_\eta \; E({\bf p})^\eta }
,
\end{equation}
where $\eta = 0$, and $Z_\eta$ is the partition function.
Adopting a reference measure of $\rho_1 =  E({\bf p})^{-1}$,
the MEP returns the modified \JUTTNER distribution,
Equation~\ref{E:RELJUTMJ} with $\eta = 1$.
\cite{DunkelET2007NJPh9} demonstrate that the choice of $\rho_0$ for
the reference measure is associated with translational invariance
in the momentum space, while $\rho_1$ is associated with 
Lorentz invariance.
As we can see, if we require a coordinate invariant form for the MEP,
we encounter the questions: Which reference measure should we use?
What is the relevant symmetry for the entropy? 
Since we do not have an established theory for relativistic kinetic theory,
the answer is not trivial.

Different approaches to relativistic kinetic theory resulted in 
particle equilibrium momentum distributions with different values of $\eta$ 
in Equation~\ref{E:RELJUTMJ}.
An extension of special relativity was also introduced along the lines 
suggested by Stueckelberg \citep{HorwitzPiron1973}. 
The problem of simultaneity is solved by establishing a measure of time 
common to all particles, on the expense of reinterpretation of relativity.
This approach resulted in an equilibrium momentum distribution of the
form of Equation~\ref{E:RELJUTMJ} with $\eta = 1$ \citep{HorwitzET1981}.
A different approach was followed by \cite{Lehmann2006JMP47}.
Lehmann noted that the derivation of \JUTTNER was relativistic, 
but not covariant. The covariant approach introduced by
Lehmann based on Poincar\'e-invariant constrained Hamiltonian dynamics
led to a relativistic one particle momentum distribution 
function of the form of Equation~\ref{E:RELJUTMJ}, with $\eta = 2$.
\cite{DunkHang2007PhyA374} were using microscopic collision processes
to investigate the relativistic generalization of the Brownian motion.
They derived an equilibrium momentum distribution for relativistic particles of
the form of the modified \JUTTNER distribution with $\eta = 1$
(Equation~\ref{E:RELJUTMJ}).

\section{Non-relativistic Sunyaev-Zel'dovich effect}
\label{S:NORELSZ}

Photons of the cosmic microwave background (CMB), 
on average, gain energy via inverse Compton 
scattering off electrons in the ICG in clusters of galaxies,
and redistributed to higher frequencies.
The intensity change in the CMB due to inverse Compton scattering in 
clusters assuming non-relativistic (Maxwellian) electron 
velocity distribution is called non-relativistic SZ effect.
Conventionally, the non-relativistic SZ effect is derived using the Kompaneets 
approximation \citep{SZ1980ARAA18}.
In the literature, the SZ amplitude derived from the Kompaneets 
approximation is identified with the non-relativisitc thermal SZ effect.
However, the Kompaneets approximation is based on more restrictive
assumptions, as we will see.

The Kompaneets equation is based on an expansion of the Boltzmann equation,
which describes the evolution of the photon occupation number, 
using a small parameter, $\Delta E_{phot} / (k_B T)$, where $E_{phot}$ is the 
energy change in the photons due to inverse Compton scattering, 
$T$ is the electron temperature, and $k_B$ is the Boltzmann constant \citep{Komp1957}.
The Kompaneets approximation assumes that the isotropic incoming photons 
have a thermal (Planckian) spectrum, the isotropic electron velocity distribution is Maxwellian,
and the energy change in the photons, $\Delta E_{phot}$ is small, 
$\Delta E_{phot} / (k_B T) << 1$.
This method is using the Thompson scattering cross section in the
rest frame of the CMB instead of the rest frame of the electron 
assuming that the electron velocities are small, thus the 
Lorentz transformations between the two rest frames can be ignored.

Sunyaev and Zel'dovich assumed that the photon occupation number, $n$, and $x_e$ 
are small in the ICG, thus terms proportional to $n$ and $n^2$ can be ignored 
relative to ${\partial n/\partial x_e}$. In this case the Kompaneets equation simplifies to

\begin{equation}  \label{E:KOMPSZ}
    \frac{\partial n}{\partial y} =  \frac{1}{x_e^2} 
                             \frac{\partial }{\partial x_e} \Biggl[  x_e^4 \Biggl( \frac{\partial n}{\partial x_e} 
                                                                          \Biggr) \Biggr]
,
\end{equation}
where the dimensionless frequency is $x_e = h_P \nu / (k_B T)$, and $h_P$ is the Planck constant,
the Compton-$y$ parameter is $y = \int \Theta d \tau$, where the dimensionless electron temperature 
is $\Theta = (k_B T)/(m_e c^2)$, $m_e$ is the mass of the electron, and the optical depth, $\tau$ is 
defined as $d \tau = \sigma_T n_e d \ell$, where $\sigma_T$ is Thomson scattering cross-section, 
$n_e$ is the electron number density, and $d \ell$ is the line element along the LOS.

Changing the dimensionless frequency $x_e$ to $x_\nu = h_P \nu /$ $(k_B T_{CMB})$, 
where $T_{CMB}$ is the temperature of the CMB,
does not change the right hand side of Equation~\ref{E:KOMPSZ}, 
and changing variables from $n(x_\nu, y)$, to $n( \ln x_\nu + 3y, y)$, 
Equation~\ref{E:KOMPSZ} can be transformed into a form of diffusion equation.
This diffusion equation can be solved analytically, and since the intensity is proportional to 
$x^3 n$, the intensity change can be expressed as
\begin{equation}  \label{E:ISZKOMP}
    \Delta I_K (x_\nu, T) = y\, \frac{\imath_0\, x_\nu^4 \,e^{x_\nu}}{(e^{x_\nu} - 1 )^2 } 
                     \biggl(  x_\nu { e^{x_\nu} + 1 \over e^{x_\nu} - 1 } - 4 \biggr)
, 
\end{equation}
where the Compton-$y$ parameter is $y = \int \Theta d \tau$, 
where the optical depth, $\tau$ is defined as $d \tau = \sigma_T n_e d \ell$,
where $\sigma_T$ is Thomson scattering cross-section, 
$n_e$ is the electron number density, and $d \ell$ is the line element along the LOS,
and the conversion factor, $i_0$, is 
\begin{equation}  \label{E:INULL}
   i_0 = 2 (k_B T_{CMB})^3 / (h_P c)^2
.
\end{equation}

In this approximation, at constant temperature in the line of sight (LOS), 
$\Delta I_K \sim \tau$, which shows that it is essentially a single
scattering approximation 
(also verified with Monte Carlo simulations; \citealt{MolnarBirk1999}).
In the single scattering approximation the photon and electron 
distributions are isotropic, in accordance with the original assumptions. 
Note, that single scattering, in most cases, is an adequate approximation in the ICG,
since the optical depth integrated along a LOS through a cluster is 
$\tau = \int \sigma_T n_e d \ell \, \simless\, 0.01$.

In the Kompaneets approximation, only the 
amplitude of the SZ effect depends on the electron temperature,
the shape of the effect as a function of frequency is the same. 
As a consequence, the crossover frequency, $\nu_0 = 217.7$ GHz, 
where the SZ amplitude is zero ($\Delta I_K [\nu_0] = 0$)
and the amplitude changes from decrement in the CMB to increment,
is independent of the electron temperature.

\section{Relativistic Sunyaev-Zel'dovich effect}
\label{S:RELSZ}

The Komponeets approximation has been and is widely used to calculate the SZ signal, 
since it can be easily calculated, as it provides an analytic expression 
(for reviews see \citealt{CarlstromET2002,Birkinshaw1999}).
However, it was realized that, at the high-temperature 
ICG ($\sim\,$15 keV), the gas is mildly relativistic, thus a relativistic generalization of the 
non-relativistic Maxwellian velocity distribution should be used at mm/submm wavelengths
\citep{Rephaeli1995ApJ445}.

The relativistic treatment of the inverse Compton scattering in mildly relativistic
electrons (applicable in the ICG) without assuming that the photon 
energy change is small was derived by \cite{Wright1979}.
Wright's method adopts the Thompson scattering cross section
in the rest frame of the electron instead of the rest frame of the CMB,
and uses Lorentz transformations between them.

Following Wright, we express the frequency change due to inverse Compton 
scattering between an electron with velocity $\beta = \vmath/c$ 
(where $c$ is the speed of light) as a function of the logarithm of the ratio 
of the frequencies of the scattered and input photons, 
$\nu$ and $\nu_0$, in the Lab frame, 

\begin{equation} \label{E:SDEF}
  s = \ln \biggl[ \frac{\nu}{\nu_0} \biggr] = 
                                          \ln \biggl[ \frac{1 + \beta \mu_2}{1 - \beta \mu_1} \biggr] 
,
\end{equation}
where we also expressed the ratio of the frequencies in the rest frame of the electron, 
where the photon scattered from an input cosine angle
$\mu_1 = \cos \varphi_1$ to $\mu_2 = \cos \varphi_2$ relative to the $z$ axis
($z$ is parallel to the velocity of the electron).

We use the Wright's formalism in the single scattering limit, which is appropriate
in the ICG with low optical depth.
In this approximation, the probability that a single photon scattering with an electron with 
a velocity of $\beta$ (in units of $c$) results in a frequency shift described by $s$, 
the frequency redistribution function, $P_1(s, \beta)$, 
can be expressed in the electrons's rest frame, as 

\begin{equation} \label{E:DEFP1SBETA}
  P_1(s, \beta) \, ds = \int 
                                    p_\mu(\mu_1) \phi(\mu_2 | \mu_1) \, d \mu_1 \frac{d \mu_2}{d s}  \,d s
,
\end{equation}
where $p_\mu(\mu_1)$ is the probability that the electron scatters with a photon 
having an incoming cosine angle $\mu_1$, and $\phi(\mu_2 | \mu_1)$ is the 
conditional probability of photons scattering into cosine angle $\mu_2$ 
if the initial photon direction cosine was $\mu_1$ ($\mu_1 \rightarrow \mu_2$) 
in the rest frame of the electron \citep{Chandra1960}:

\begin{equation} \label{E:FI12}
  \phi(\mu_2 | \mu_1) d \mu_2 = \frac{3}{8} 
               \Bigl[ 1 + \mu_1^2 {\mu_2}^2 + \frac{1}{2} (1 - \mu_1^2) )(1 - {\mu_2}^2) \Bigr]  d \mu_2
.
\end{equation}
$p_\mu(\mu_1)$ can be derived from transforming an isotropic incoming
photon direction distribution to the rest frame of the electron, 

\begin{equation} \label{E:PMU}
  p_\mu (\mu_1) d \mu_1 =  \bigl[  2 \gamma^4 (1 - \beta \mu_1)^3 \bigr]^{-1}   d \mu_1
,
\end{equation}
where $\gamma^2 = 1/(1 - \beta^2)$.
The frequency redistribution function due to Compton scattering,
the probability for a single scattering resulting a frequency shift $s$ in the Lab frame,
thus can be expressed as 

\begin{equation} \label{E:P1SBETAFI}
  P_1(s, \beta) d s = \frac{1}{2 \gamma^4 \beta} 
                      \int_{\mu_a}^{\mu_b} d \mu_1\,
                                                   \frac{1 + \beta \mu_2}{ [1 - \beta \mu_1]^3}  \phi(\mu_2 | \mu_1) d s
,
\end{equation}
where the limits of the integral, $\mu_a$ and $\mu_b$ are determined by the condition 
that the cosines have to be real:

\begin{eqnarray}         \label{e:mu1mu2}
   \mu_a & =  &   \begin{cases}
                                                                       -1                       &     s  \le 0    \cr
                                                   [1 - e^{-s}(1+\beta)] / \beta    &     s \ge 0   \cr
                          \end{cases} \cr
   \mu_b & =  &   \begin{cases}
                                                   [1 - e^{-s}(1-\beta) ] / \beta      &      s \le 0    \cr
                                                                      1                           &      s \ge 0.
                           \end{cases}              
\end{eqnarray}
The integral over $\mu_1$ can be performed analytically (e.g., \citealt{Molnar2015}).

The frequency redistribution function, $P_1$, as a function of $s$, can be obtained by 
integrating over the velocity distribution of the electrons, $p_e (\beta) d \beta$,
assuming that it is known, 
\begin{equation} \label{E:P1S}
  P_1(s) = \int_{\beta_0}^1 P_1(s, \beta)  p_e (\beta) \,d\beta
,
\end{equation}
where the lower limit $\beta_0$ is the minimum electron velocity needed to get frequency shift $s$, 

\begin{equation}    
   \beta_0 =  {e^{|s|} - 1 \over e^{|s|} + 1} 
.
\end{equation}
Note that carrying the integral over $\mu_1$ analytically, our integral in Equation~\ref{E:P1S}
is only one dimensional. In general, this integral needs to be performed numerically.

The intensity at frequency $\nu$, after single scatterings, $I_1 (\nu)$, can be expressed 
as a convolution of the incoming intensity, $I_0(\nu_0)$, assumed to be isotropic,
and the single scattering probability distribution, $P_1(s)$, 

\begin{equation}  \label{E:I1ALL}
   I_1 (\nu) = \int_{-\infty}^{\infty} d s P_1(s) I_0 (\nu_0) 
,
\end{equation}
where the integral is over $s = \ln (\nu / \nu_0)$ with the frequency after scattering, $\nu$, fixed,
and $P_1(s)$ is given by Equation~\ref{E:P1S}.

In the single-scattering approximation, 
the probability that a photon passes the ICG without scattering
is $e^{\tau}$, and the chance that is scatted once is $\tau e^{\tau}$.
Thus the full frequency redistribution function, 
which describes the frequency change after the photons pass through the ICG, becomes
\begin{equation}  \label{E:F1ALL}
   F_1 (s) = (1 - \tau) \delta(s) + \tau \,P_1(s)
,
\end{equation}
and the emergent intensity can be expressed as a convolution, 

\begin{equation}  \label{E:I1SZALL}
   I_{SZ} (\nu) = \int_{-\infty}^{\infty} d s F_1(s) I_0 (\nu_0) 
.
\end{equation}
Thus, assuming that the incoming photon frequency distribution is Planckian in the Lab frame,
the change in the emerging intensity, $I_{SZ} (\nu) - I_0 (\nu)$, the thermal SZ effect, becomes

\begin{equation}  \label{E:DELSZ}
   \Delta I_{SZ} (\nu) =  i_0 \tau  \int d s \,P_1(s) 
                                \left( \frac{x_0^3}{e^{x_0} - 1} - \frac{x_\nu^3}{e^{x_\nu} - 1} \right)
,
\end{equation}
where $x_0 = h_P \nu_0 / (k_B T_{CMB})$, and $i_0$ is defined as in Equation~\ref{E:INULL}.

Assuming that the electron equilibrium velocity distribution function 
is non-relativistic (Maxwellian), 
\begin{equation}
      p_{MX} (\beta) \, d \beta = N_{MX} \beta^2 \, \exp \bigl\{ - \beta^2 / (2 \Theta)  \bigr\} \, d \beta
,
\label{E:PELMAXW}
\end{equation}
%
where $N_{MX}$ is the normalization, we obtain a non-relativistic approximation of the 
SZ effect, $\Delta I_{NR}$ from Equation~\ref{E:DELSZ}.
However, contrary to the conventional non-relativistic SZ effect based on the Kompaneets
approximation, this method uses the Wright formalism, and 
does not ignore the Lorentz transformations 
between the rest frame of the electron and the CMB.

\cite{Wright1979} calculated the SZ amplitude using the relativistic 
generalization of the Maxwellian velocity distribution derived by \JUTTNER,
\begin{equation}
      p_{J} (\beta) \, d \beta = N_J \frac{\gamma^5 \beta^2}{\Theta}
                                                             \exp \bigl\{ - \gamma / \Theta  \bigr\} \, d \beta
,
\label{E:PELJUT}
\end{equation}
%
where $N_J$ is the normalization for the probability distribution.
The relativistic SZ effect based on the \JUTTNER velocity distribution, 
$\Delta I_{J}$, can be derived using Equation~\ref{E:PELJUT} in Equation~\ref{E:DELSZ}.

We use the modified \JUTTNER distribution in the form of
\begin{equation}
      p_{MJ\eta} (\beta) \, d \beta = N_{MJ\eta} \frac{\gamma^5 \beta^2}{\gamma^\eta \, \Theta}
                                                             \exp \bigl\{ - \gamma / \Theta  \bigr\} \, d \beta
,
\label{E:PELMJUT}
\end{equation}
%
where $N_{MJ\eta}$ is the normalization for the modified \JUTTNER electron velocity distribution
\citep{Lehmann2006JMP47,DunkHang2007PhyA374}.
The modified \JUTTNER distribution differs from
the \JUTTNER distribution only in the power of $\gamma$, $p_{MJ\eta} \sim p_{J}/\gamma^\eta$
(e.g., \citealt{DunkelET2007NJPh9}).
We derive the relativistic SZ effect based on the modified \JUTTNER velocity distribution,
$\Delta I_{MJ\eta}$, using Equation~\ref{E:DELSZ}.

\begin{figure}[t]
\includegraphics[width=.47\textwidth]{\FIGURES/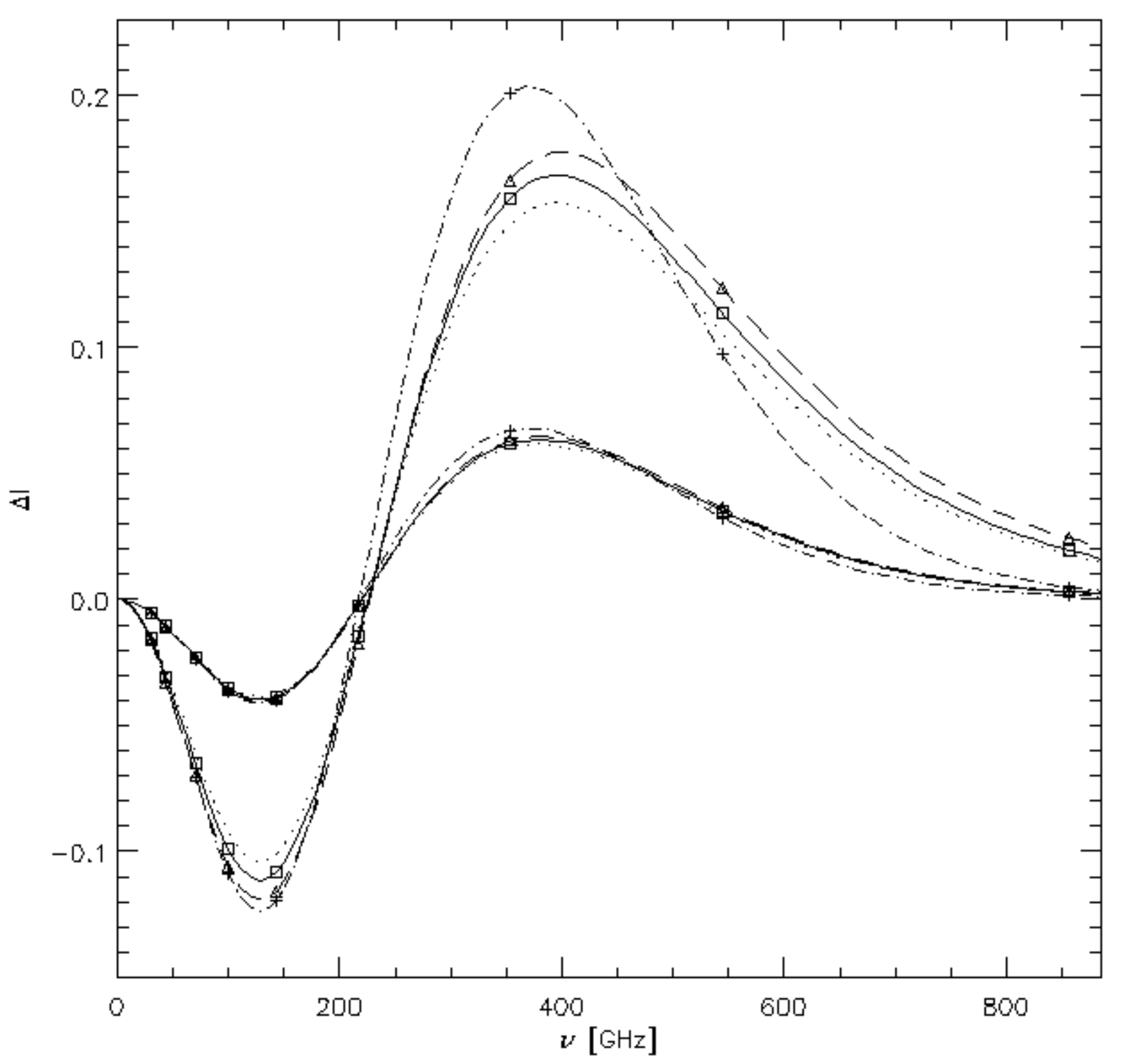}
\caption{
The Sunyaev-Zel'dovich effect (in units of $i_0 \tau$)
as a function of frequency (in GHz) for electron temperatures of
15.33 keV and 5.11 keV (upper and lower set of lines).
Solid, dotted, and dashed lines represent SZ amplitudes 
assuming electron velocity distribution functions of the form of
relativistic \JUTTNER, modified \JUTTNER \textcolor{magenta}{with $\eta = 1$}, and non-relativistic Maxwellian 
using the Wright formalism.
Dash-dotted lines show the non-relativistic SZ amplitude derived using
the Kompaneets approximation.
The squares, triangles, and plus signs represent the corresponding SZ amplitudes 
at the frequencies of the \PLANCK instruments.
\vspace{0.2 cm}
\label{F:NUTOSZ}
}
\end{figure}

As an illustration, in Figure~\ref{F:NUTOSZ}, 
we show the intensity change in the CMB  due to the Sunyaev-Zel'dovich effect 
(in units of $i_0 \tau$, where $\tau$ is the optical depth and $i_0$ is 
given by Equation~\ref{E:INULL})
as a function of frequency (in GHz) for electron temperatures of
15.33 keV and 5.11 keV (upper and lower set of lines).
Solid, dashed, and dotted lines represent SZ amplitudes 
assuming electron velocity distribution functions of the form of
relativistic \JUTTNER (Equation~\ref{E:PELJUT}), 
modified \JUTTNER with $\eta = 1$ (Equation~\ref{E:PELMJUT}), 
and non-relativistic, Maxwellian (Equation~\ref{E:PELMAXW})
using the Wright formalism 
($\Delta I_{J}$, $\Delta I_{MJ1}$, and $\Delta I_{MX}$).
Dash-dotted lines show the non-relativistic SZ amplitude derived based on 
the Kompaneets approximation ($\Delta I_{K}$).
The squares, triangles, and plus signs represent SZ amplitudes assuming 
relativistic \JUTTNER and Maxwellian velocity distributions based on 
the Wright formalism, and amplitudes using 
the Kompaneets approximation at the frequencies of the \PLANCK instruments
($\nu =$ 30, 44, 70, 100, 143, 217, 353, 545, and 857 GHz; 
e.g., \citealt{BourdinET2017ApJ843}; note, the \HERSCH-SPIRE instrument
also covers the 600 and 857 GHz frequency channels with higher spatial resolution;
\citealt{GriffinET2010AA518}).

%
%

Figure~\ref{F:NUTOSZ} demonstrates that the differences between 
SZ amplitudes derived from the Kompaneets approximation (dash-dotted lines)
and those derived from the Wright formalism using Maxwellian velocity distribution
(dashed lines) relative to those derived from assuming \JUTTNER distribution (solid lines)
are larger 
(e.g., $\Delta I_{K}$ is much larger than $\Delta I_{J}$ or $\Delta I_{MX}$ at 350 GHz).
As a consequence, we obtain a better non-relativistic approximation for 
the SZ effect using the Wright method assuming a Maxwellian velocity distribution
as opposed to using the Kompaneets equation.
Thus, the main difference between the amplitudes of the SZ effect 
based on the Kompaneets approximation and the relativistic SZ effect
using the Wright method
is not the form of the velocity distribution used (Maxwelian vs. \JUTTNER),
but the treatment of the collision process
(compare dashed and dash-dotted lines to the solid lines in Figure~\ref{F:NUTOSZ}).
The komponeets approximation over(under) estimates the SZ signal 
at frequencies $\simless 450$ GHz ($\simgreat 450$ GHz).
In Table~\ref{T:TABLE1} we quantify these differences.
In this table we show the ratios between SZ amplitudes
derived from the Kompaneets approximation over
those based on the Wright formalism assuming \JUTTNER distribution, 
$\Delta I_{K}/\Delta I_{J}$, and Maxwellian distribution, $\Delta I_{K}/\Delta I_{MX}$, 
at electron temperatures of $T = 15.55$ keV and $T = 5.11$ keV.

At high electron temperatures, e.g., $T = 15$ keV, 
the error due to using the Kompaneets approximation instead of the 
Wright formalism with the \JUTTNER (or Maxwellian) distribution 
is $\simgreat\,$10\% for $\nu \, \simgreat\, 100$ GHz (or $\nu \,\simgreat\, 217$ GHz).
At the highest \PLANCK frequency, $\nu = 857$ GHz, $\Delta I_{K}$ and $\Delta I_{MX}$ 
are lower than $\Delta I_{J}$ by a factor of 4 and 5 
($\Delta I_{K}/\Delta I_{J} = 0.25$, $\Delta I_{K}/\Delta I_{MX} = 0.2$).
Even at lower temperatures, $\Delta I_{K}$ deviates from $\Delta I_{J}$
by $\simgreat\,$8\%, and at $\nu = 857$ GHz $\Delta I_{K}$ is a factor of 2 too small
($\Delta I_{K}/\Delta I_{J} \sim \Delta I_{K}/\Delta I_{MX} \sim 0.5$).
Thus, Table~\ref{T:TABLE1} demonstrates that at high \PLANCK frequencies
($\nu \,\simgreat\, 217$ GHz), even for lower ICG temperatures ($T \sim 5$ keV),
the Wright formalism should be used to calculate the SZ signal
either with the \JUTTNER, or Maxwellian velocity distribution instead 
of the Kompaneets approximation.
However, the Kompaneets approximation is still widely used at all \PLANCK 
frequencies (e.g., \citealt{BaldiET2019AA630,PLANCKXXII2016}).

%
%
\begin{deluxetable}{rcccc}[t]
\tablecolumns{4}
\tablecaption{                       \label{T:TABLE1} 
 SZ amplitude ratios at \PLANCK frequencies using different methods for $T =$ 15.55 keV and 5.11 keV. \\
} 
\tablewidth{0pt} 
\tablehead{ 
 \multicolumn{1}{c}   {Freq.}                                                   &
 \multicolumn{1}{c}   {$\Delta I_{K}/\Delta I_{J}$\,\tablenotemark{a}}      &
 \multicolumn{1}{c}   {$\Delta I_{K}/\Delta I_{MX}$\,\tablenotemark{b}}  &
 \multicolumn{1}{c}   {$\Delta I_{K}/\Delta I_{J}$\,\tablenotemark{c}}      &
 \multicolumn{1}{c}   {$\Delta I_{K}/\Delta I_{MX}$\,\tablenotemark{d}} 
}
\startdata  
    GHz$\;$      &   $T$ = 15.5 keV    &    $T$ = 15.5 keV  &  $T$ = 5.1 keV  &   $T$ = 5.1 keV   \\ \hline
    30      &   1.06    &   0.98     &   1.02     &   0.99   \\ \hline
    44      &   1.06    &   0.99     &   1.02    &  1.00  \\ \hline
    70      &   1.08    &   1.01     &   1.03    &  1.00.   \\ \hline
  100      &   1.10    &   0.03     &   1.03    &   1.01   \\ \hline
  143      &   1.11    &   1.03     &   1.04    &   1.01   \\ \hline
  217      &   0.11    &   0.09     &   0.24    &   0.22   \\ \hline
  353      &   1.26    &   1.20    &   1.08    &   1.06   \\ \hline
  545      &   0.86    &   0.79     &   0.92    &   0.89   \\ \hline
  857      &   0.25    &   0.20     &   0.53    &  0.50   
\enddata
\tablenotetext{a}{SZ amplitudes derived from the Kompaneets approximation over
                           those based on the Wright formalism assuming \JUTTNER velocity distribution
                           at $T = 15.55$ keV.}
\tablenotetext{b}{SZ amplitudes derived from the Kompaneets approximation over
                           those based on the Wright formalism assuming Maxwellian velocity distribution
                           at $T = 15.55$ keV.}
\tablenotetext{c}{Same as $a$ but for $T = 5.11$ keV.}
\tablenotetext{d}{Same as $b$ but for $T = 5.11$ keV.\\
}
\end{deluxetable}  

In contrast to the non-relativistic SZ effect (based on the Kompaneets
approximation), the shape of the relativistic SZ effect 
(derived using the Wright formalism) as a function
of frequency does depend on the temperature as well.
This make it possible to use the frequency dependence of
the SZ effect to derive the temperature of the ICG
(e.g., \citealt{PointET1998AA336,HansenET2002ApJ573}).
Other methods also have been proposed to derive the temperature 
in the ICG using the frequency dependence of the relativistic SZ effect based on 
the shift of the crossover frequency from $\nu_0 = 217.7$ GHz \citep{Rephaeli1995ApJ445},
the slope of the SZ effect near the crossover frequency \citep{ColaET2009AA494}, and 
the ratio of the SZ intensities at two different frequencies \citep{ProkET2010AA524}.

\section{Fitting to SZ observations of the Bullet cluster}
\label{S:SZBULLET}

The Bullet cluster (1E0657--56) is one of the few high-infall velocity 
merging galaxy clusters, which provided the first direct evidence for 
the existence of dark matter based on multi-frequency observations \citep{CloweET2006ApJ648}.
The offset between the mass surface density centers derived from gravitational lensing
and the X-ray emission peaks marking the gas (baryonic) component 
were significant (200--300 kpc, e.g., \citealt{ParaficzET2016AA594}).

\begin{figure}[t]
\includegraphics[width=.47\textwidth]{\FIGURES/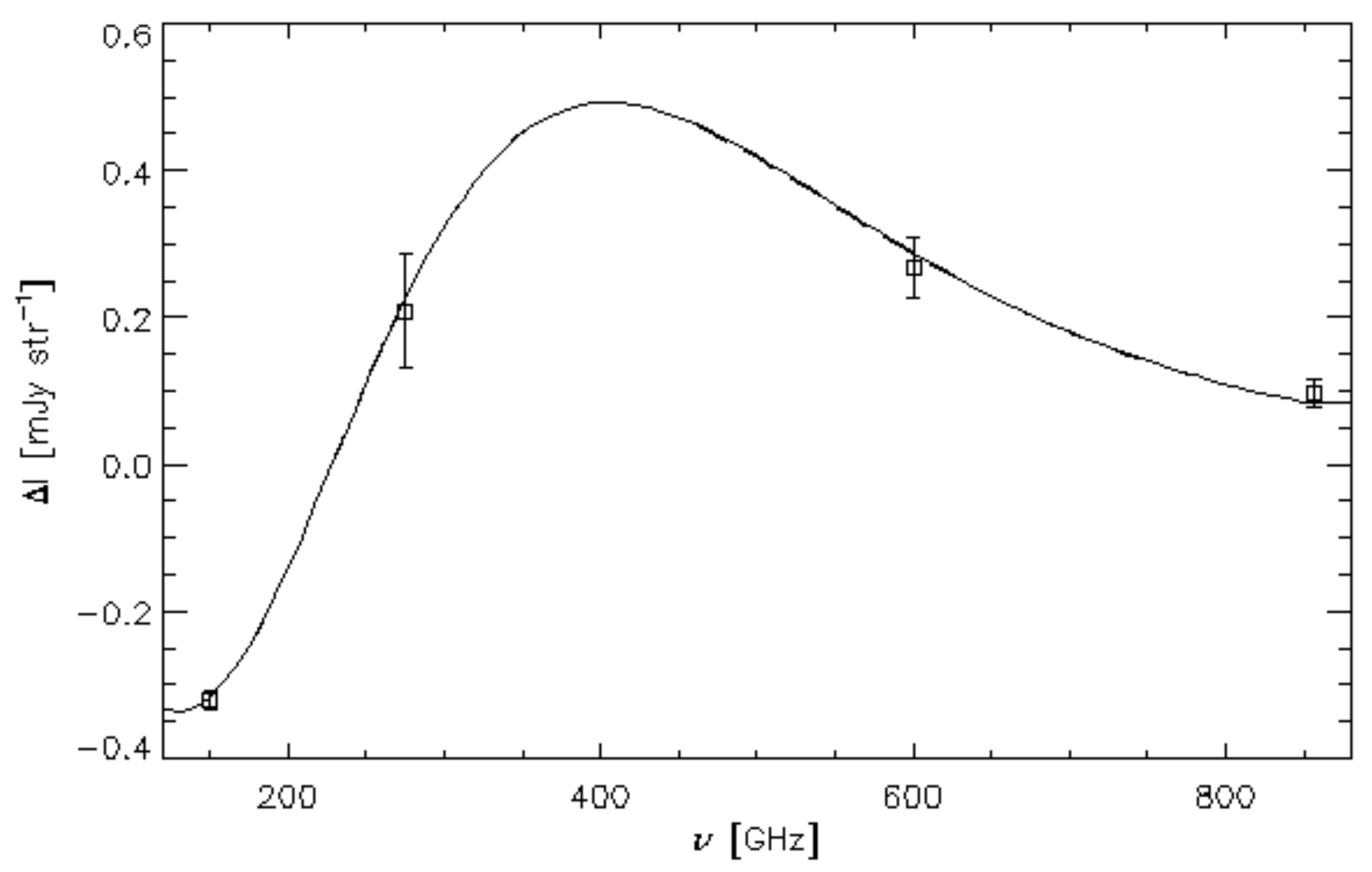}
\caption{
Sunyaev-Zel'dovich observations of the Bullet cluster (squares with error bars)
as a function of frequency and best-fit models based on different
electron velocity distributions (lines).
The solid, dash-dotted, dashed, and dash-dot-dot-dotted lines are the best-fit models 
assuming relativistic (\JUTTNER), modified \JUTTNER with $\eta = 1$ and $\eta = 1$,
and non-relativistic (Maxwellian) velocity distributions.
Note that SZ models based on all of these distributions provide a good fit, 
thus the fitted models (lines) are indistinguishable.
\vspace{0.1 cm}
\label{F:FITSZBULLET}
}
\end{figure}

We illustrate our proposed method to determine the EV distribution function
in the ICG in clusters of galaxies based on the frequency dependence 
of the SZ effect using SZ observations of the Bullet cluster.
We use archival SZ observations of the Bullet cluster available 
at four frequencies: 150, 275, 600, and 857 GHz
\citep{Gomez2004,HalversonET2009,PlaggeET2010,ZemcovET2010AA518}.
We show the SZ observations as a function of frequency in Figure~\ref{F:FITSZBULLET}
(squares with error bars).
We assume that the optical depth is unknown, and treat it as a nuisance parameter, 
therefore our method is sensitive only to the shape of the SZ signal. 
Note, that if the optical depth were known with high precision, 
the SZ amplitudes could also be used and it would be easier 
to distinguish between different velocity distributions (see Figure~\ref{F:NUTOSZ}).
We adopt a $\chi^2$ statistic and maximize our likelihood function, ${\cal L} \sim exp(-\chi^2)$
to determine the best-fit temperatures using different SZ effect models derived 
based on the Wright formalism (Section~\ref{S:RELSZ}) assuming 
relativistic \JUTTNER (Equation~\ref{E:PELJUT}), 
modified \JUTTNER with $\eta = 1$ and 2 (Equation~\ref{E:PELMJUT}), 
and non-relativistic, Maxwellian (Equation~\ref{E:PELMAXW}),
velocity distribution functions. We do not consider models derived from the 
Kompaneets equation because it does not provide a good approximation
for the nonrelativistic SZ effect at high frequencies (see Figure~\ref{F:NUTOSZ}).

%
%

We show our results in Figure~\ref{F:FITSZBULLET}.
The solid, dashed, dash-dotted, and dash-dot-dot-dotted, lines are the best-fit models assuming 
relativistic (\JUTTNER), modified \JUTTNER distributions with $\eta = 1$ and 2, 
and non-relativistic (Maxwellian) EV distributions.
The SZ model based on each one of these distributions provides a good fit, 
there is no significant difference in their $\chi^2$ values ($\Delta \chi^2 < 1$).
thus the fitted models (lines) are indistinguishable.
We obtain good fits for all four models with best-fit electron temperatures of
$T_{J} = 22.1_{-4.97}^{+5.65}$ keV, 
$T_{MJ1} = 23.0_{-5.32}^{+6.16}$ keV,  
$T_{MJ2} = 23.9_{-5.57}^{+6.34}$ keV,   
and $T_{MX} = 18.3_{-3.76}^{+4.33}$ keV, 
assuming \JUTTNER, modified \JUTTNER with $\eta = 1$ and $\eta = 2$, 
and Maxwellian EV distributions.
Our result for the best-fit temperature, $T_{J} = 22$ keV, assuming the \JUTTNER EV 
distribution agrees with that obtained by \cite{Cola2011AA527} 
assuming a single temperature gas and adopting the same velocity distribution function.

\section{Fitting different velocity distribution functions to simulated SZ observations}
\label{S:SZMC}

We address the question:
What accuracy do we need in the SZ observations to be 
able to derive more accurate temperatures and thus
distinguish EV distribution functions with high significance?
We proceed in two steps: 
First, we carry out Monte Carlo simulations to generate mock SZ observations
assuming a fiducial EV distribution based on the \JUTTNER 
distribution function assuming a few per cent errors in the SZ measurements.
Then we determine the best-fit temperatures 
(applying $\chi^2$ statistic as in Section~\ref{S:SZBULLET})
using data from the same Monte Carlo realization of 
the SZ observations assuming different EV distribution functions, 
and compare the probability distributions of the temperatures.
We are adopting modified \JUTTNER distributions only with $\eta = 1$ and 2,
since these are suggested by different approaches of relativistic kinetic theory
(as described in Section~\ref{S:RELKIN}). We found no theory resulting
modified \JUTTNER distributions with $\eta > 2$.
The distributions of the best-fit values can be used to determine the errors in the derived 
gas temperatures for the assumed per cent errors in the SZ amplitude measurement.
Essentially, we use Monte Carlo simulations to propagate the errors
from the SZ amplitude measurements to the derived temperatures.

%
%
\begin{figure}[t]
\includegraphics[width=.47\textwidth]{\FIGURES/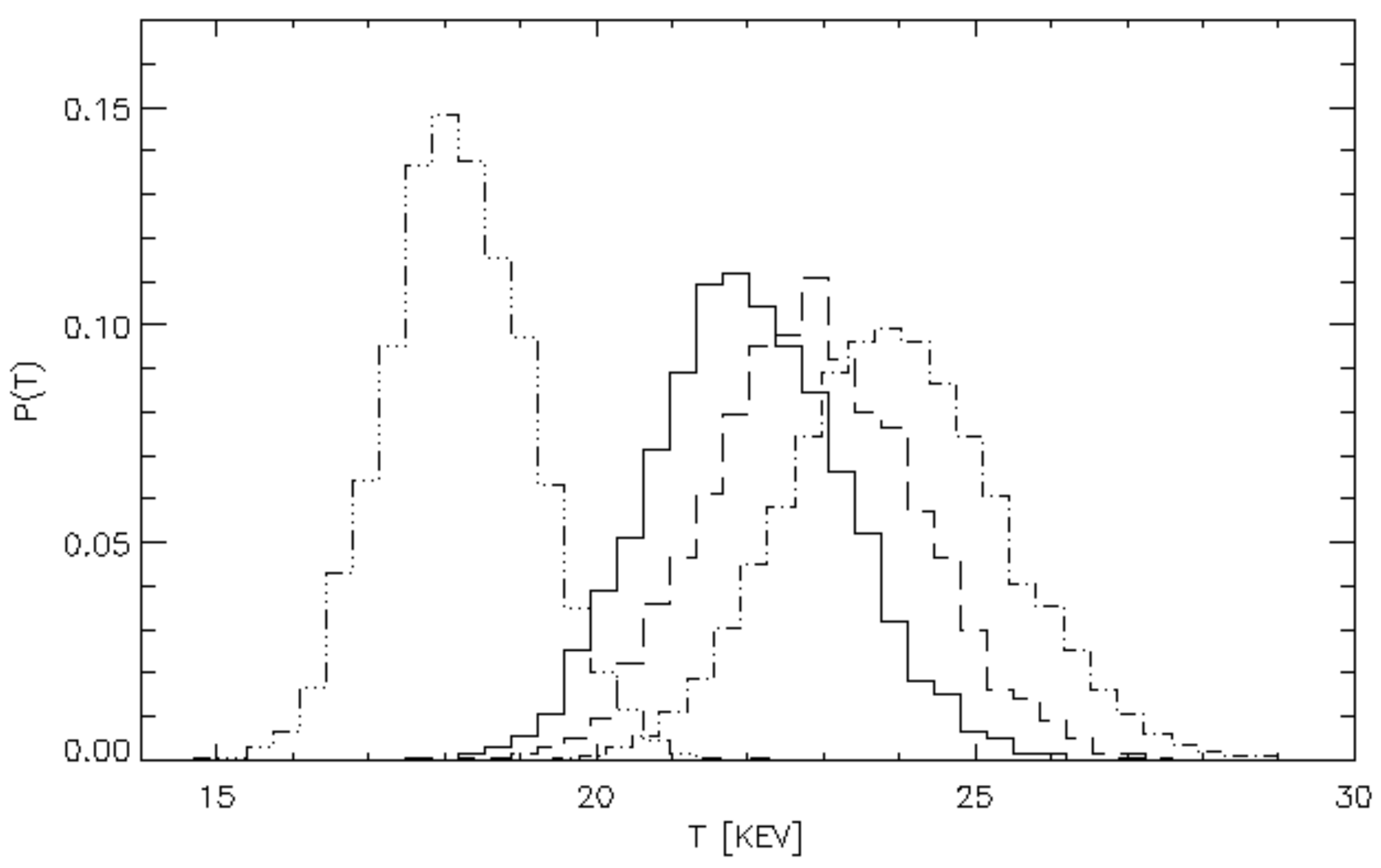}
\caption{
Probability distributions of best-fit temperatures
from fitting SZ models based on the \JUTTNER, modified \JUTTNER 
with $\eta = 1$ and 2, and Maxwellian electron velocity distributions 
to the same simulated SZ observations 
(solid, dashed, and dash-dotted, and dash-dot-dot-dotted lines).
We assumed a \JUTTNER velocity distribution as our fiducial model 
with a temperature of 22.1 keV, and 5\% errors in the SZ measurements.
\label{F:MC5PER}
}
\end{figure} 

\subsection{Fitting to SZ observations of a single-temperature gas}
\label{SS:FITPLASMA}

In this section we assume a single temperature gas. For definiteness, motivated 
by our fit to the Bullet cluster data (Section~\ref{S:SZBULLET}),
we adopt an electron temperature of 22.1 keV.

We display our results in Figure~\ref{F:MC5PER} assuming 5\% error in the SZ measurements. 
In this figure we show probability distributions of the best-fit temperatures 
from fitting SZ models based on the \JUTTNER, $P_{J}(T)$, 
modified \JUTTNER with $\eta = 1$ and $\eta = 1$, $P_{MJ1}(T)$ and $P_{MJ2}(T)$,
and Maxwellian, $P_{MX}(T)$, EV distributions to the same simulated 
SZ observations with solid, dashed, dash-dotted, and dash-dot-dot-dotted lines.
As before, we find that each SZ model provides a good fit to the simulated data,
and there is no significant difference between the fitted $\chi^2$ values.
The shapes of the fitted functions are very similar, they differ less than a fraction of 1\%
(the best-fit SZ models not shown, they would be indistinguishable, as in 
Figure~\ref{F:FITSZBULLET}).
This agrees with the result of \cite{ProkhorovET2011AA529},
who demonstrated that their method needs an accuracy of 0.1\%
in the SZ measurements to be able to distinguish between 
Maxwellian and \JUTTNER EV distributions.

Even though the best-fit SZ models based on different EV distributions are 
indistinguishable, we find, that the fitted temperatures are different: 
we obtained best-fit electron temperatures of 
$T_{J} = 22.1_{-1.25}^{+1.25}$ keV, 
$T_{MJ1} = 23.0_{-1.41}^{+1.41}$ keV, and  
$T_{MJ2} = 23.9_{-1.43}^{+1.43}$ keV, and  
$T_{MX} = 18.3_{-0.96}^{+0.96}$ keV 
for EV distributions of the form of \JUTTNER, modified \JUTTNER
with $\eta = 1$ and 2, and Maxwellian.
The probability distributions for the best-fit temperatures assuming
Maxwellian EV distribution ($P_{MX}[T]$) is well separated from those based on 
its three relativistic generalizations, the \JUTTNER and modified \JUTTNER 
velocity distributions with $\eta = 1$ and 2
($P_{J}[T]$, $P_{MJ1}[T]$, and $P_{MJ2}[T]$; Figure~\ref{F:MC5PER}), 
suggesting that the Maxwellian EV distribution function can be distinguished from 
the other three distributions.
However, our results indicate that the \JUTTNER and modified \JUTTNER 
EV distribution functions cannot be distinguished based on their temperatures 
assuming 5\% accuracy in the SZ measurements.

\begin{figure}[t]
\includegraphics[width=.47\textwidth]{\FIGURES/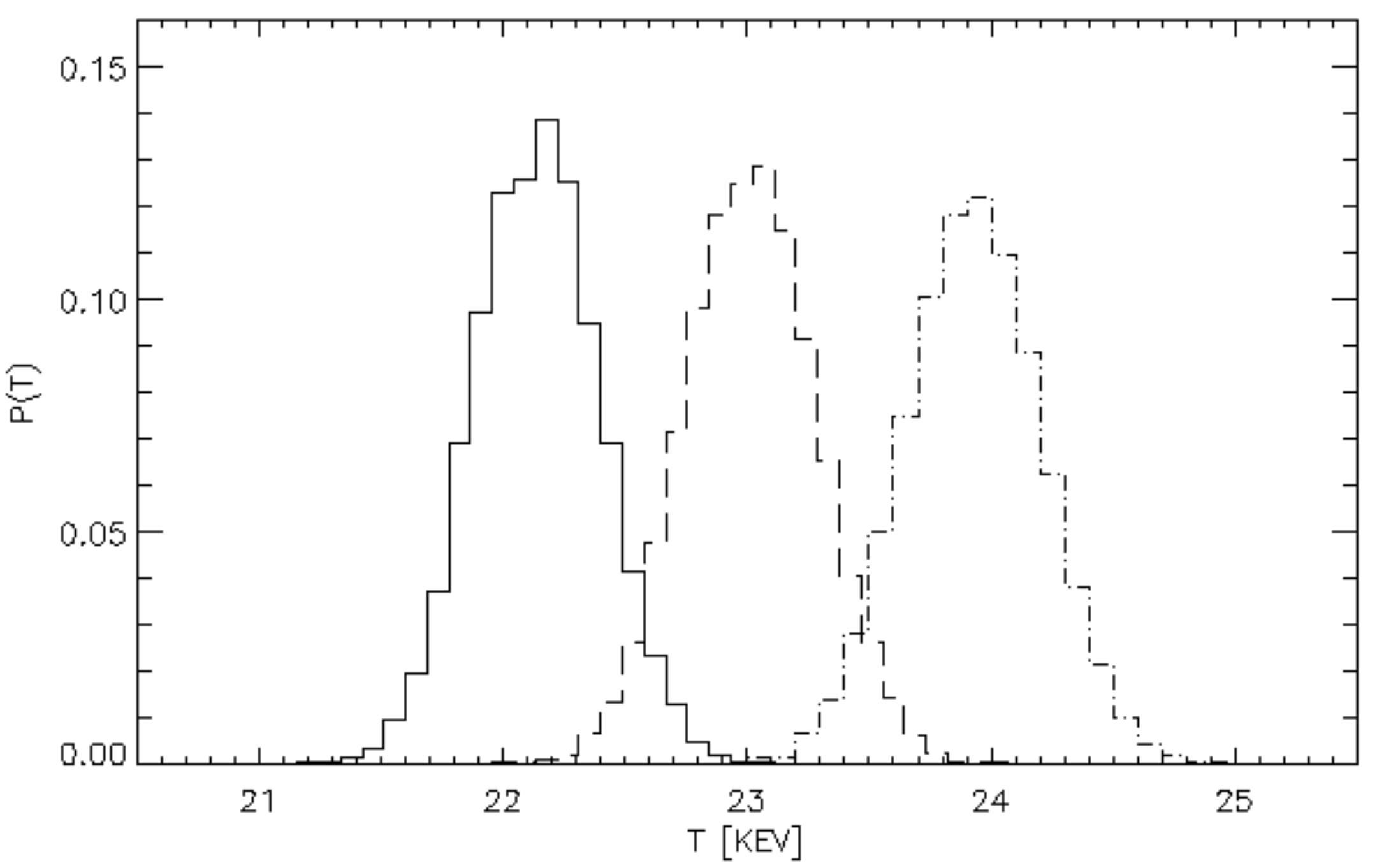}
\caption{
Same as Figure~\ref{F:MC5PER}, assuming the same fiducial model 
and fitting SZ models based on the 
\JUTTNER and modified \JUTTNER electron velocity distributions with $\eta = 1$ and 2 
(solid, dashed, and dash-dotted lines), but adopting measurement errors of 1\%
in the SZ effect.
\vspace{0.3 cm}
\label{F:MC1PER}
}
\end{figure}

We carried out Monte Carlo simulations to estimate the accuracy 
necessary to distinguish between the different relativistic velocity distributions:
the \JUTTNER and modified \JUTTNER distributions. 
We used the same fiducial model as before: \JUTTNER EV distribution function 
with a temperature of 22.1 keV, but this time we assumed 1\% measurement error 
in the SZ observations.
The probability distributions of best-fit temperatures 
from fitting SZ models based on the \JUTTNER, $P_{J}(T)$, and 
modified \JUTTNER electron velocity distributions with $\eta = 1$ and 2,
$P_{MJ1}(T)$ and $P_{MJ2}(T)$ to the same simulated SZ observations 
(solid, dashed, and dash-dotted lines) are shown in Figure~\ref{F:MC1PER}.

Again, we find that all models provide a good fit to the simulated
data, the shape of the fitted SZ models are very similar.
However, the fitted temperatures are different: 
we obtain a good fit for all three models with best-fit electron temperatures: 
$T_{J} = 22.1_{-0.25}^{+0.25}$ keV, 
$T_{MJ1} = 23.0_{-0.27}^{+0.27}$ keV   
$T_{MJ2} = 23.9_{-0.28}^{+0.28}$ keV
assuming \JUTTNER and modified \JUTTNER EV distributions with $\eta = 1$ and 2.
Note that the best-fit temperatures are the same for these two EV
distribution functions as before, since we used the same fiducial models.
The differences are only in the errors in the derived temperatures, 
which are much smaller as a consequence of our adopted smaller
errors in the SZ amplitudes. 
The probability distributions for the best-fit temperatures assuming
\JUTTNER ($P_{J}[T]$) and modified \JUTTNER $P_{MJ1}[T]$, $P_{MJ2}[T]$) 
velocity distributions 
are well-separated (Figure~\ref{F:MC1PER}), suggesting that a 1\% error in the 
SZ measurements would allows us to distinguish between them.

%
%

\subsection{Fitting to SZ observations of a shock from a merging cluster simulation}
\label{SS:FITSHOCK}

In the previous section we demonstrated that, in the case of a single
temperature gas, SZ measurements with an accuracy of 1\% at
four frequencies may be used to distinguish between EV distribution 
functions of the form of non-relativistic Maxwellian, \JUTTNER, and modified \JUTTNER.
However, any LOS through a galaxy cluster contains a range of temperatures, 
even if the cluster is in dynamical equilibrium.
It is also difficult to model clusters because they are not spherical, 
may be dynamically active, and contain substructure.

In this section we use our \NBODYHYD simulation of a merging 
galaxy cluster to test our method to distinguish between EV distribution
functions. Our simulation was carried out using \FLASH,
an Eulerian \NBODYHYD code developed at the 
Center for Astrophysical Thermonuclear Flashes at the University of Chicago
\citep{Fryxell2000ApJS131p273,Ricker2008ApJS176}.
We used our well-tested method to setup and run the simulation 
\citep[e.g.,][]{MolnarBroadhurst2015,MolnarBroadhurst2017,MolnarBroadhurst2018}.
For a detailed description of our method, see \cite{MolnarET2012ApJ748}.  
We adopted initial total masses of 
$1.7 \times 10^{15} \,M_\odot$ and $1.6 \times 10^{15} \,M_\odot$ 
(main and infalling cluster), an impact parameter of 100\,kpc, 
and an infall velocity of $2500\,$\KMSEC in our simulation.

%
%
\begin{figure}[t]
\includegraphics[width=.46\textwidth]{\FIGURES/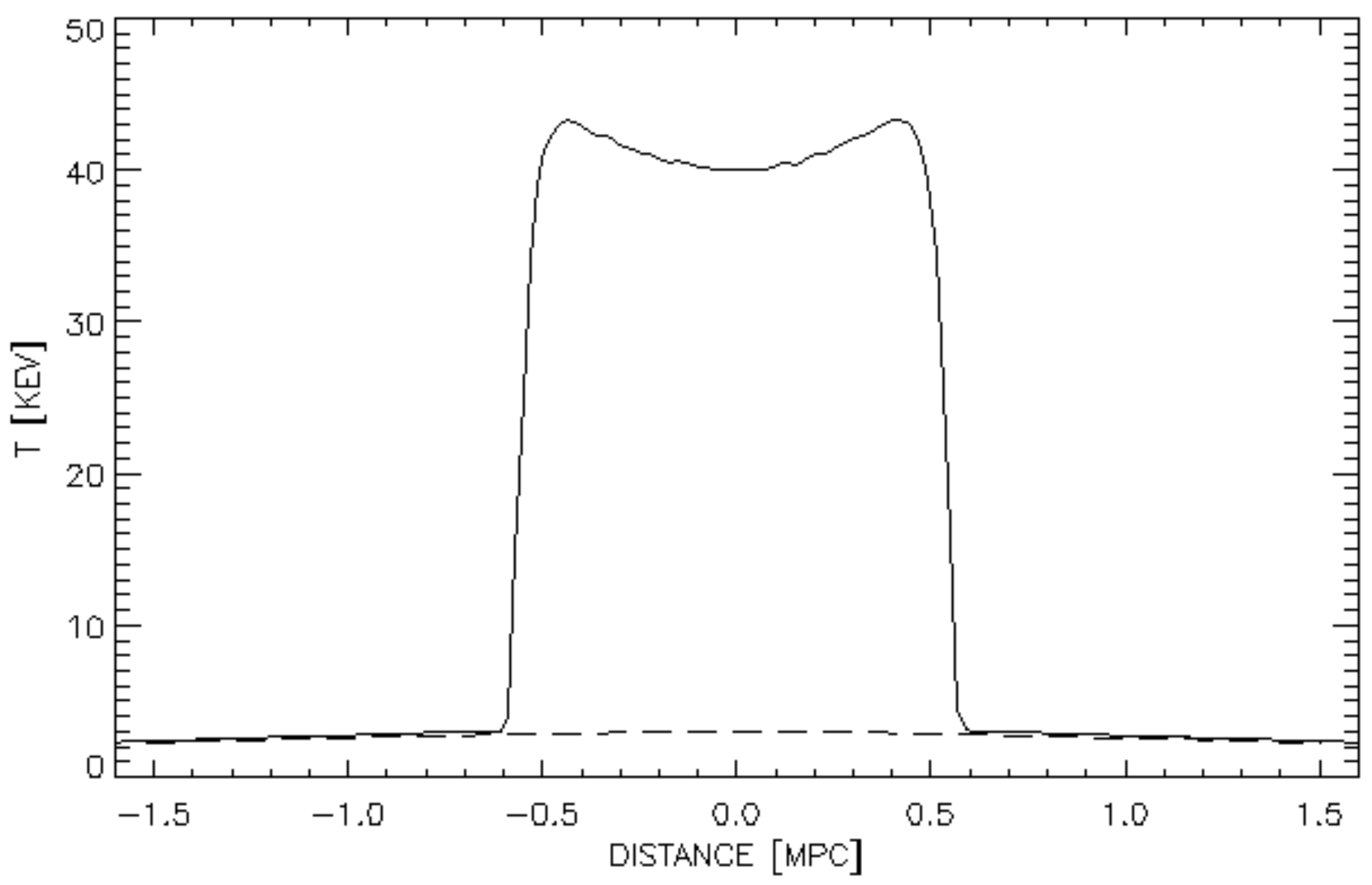}
\caption{
Temperature distribution along the LOS through a shock
and a pre-shocked region near the shock extracted from a
merging cluster \NBODYHYD simulation
(solid and dashed lines; see text for details).
\label{F:TKEVSHOCK}
}
\end{figure} 

We chose an epoch soon after the 1st core passage from our outputs, 
when a bow shock is moving ahead of the infalling cluster. 
The Mach number for this shock is 6.51, and the shock velocity is 5740 \KMSEC.
We extract data using a viewing angle assuming that 
the two cluster centers and their relative velocities 
are in the plane of the sky.
We choose a LOS close to the edge of the bow shock.
We show the temperature distribution along this LOS in 
Figure~\ref{F:TKEVSHOCK} (solid line).
The  temperature along a LOS through the pre-shocked gas near the shock is 
shown with a dashed line.
The average temperature through the LOS of the pre-shocked gas
is 2.68 keV with a 8.5\% dispersion, while the average temperature
of the pre-shock region in the LOS through the shocked gas 
(low temperature region of the solid line in Figure~\ref{F:TKEVSHOCK}) 
is 2.64 keV with a 8.8\% dispersion. 
The average temperature in the shocked region is 42.1 keV with a 2\%
dispersion.
Thus, we can identify two phases of the gas in the LOS 
through the shock: one lower and one with temperature phase corresponding
to the pre-shocked and shocked regions, with a less than 9\% dispersion,
much less than the difference between the average temperatures (42.1 keV vs. 2.68 keV).
These results suggest that we may adopt a two temperature model 
for the gas in the LOS through the shock.

\begin{figure}[t]
\includegraphics[width=.477\textwidth]{\FIGURES/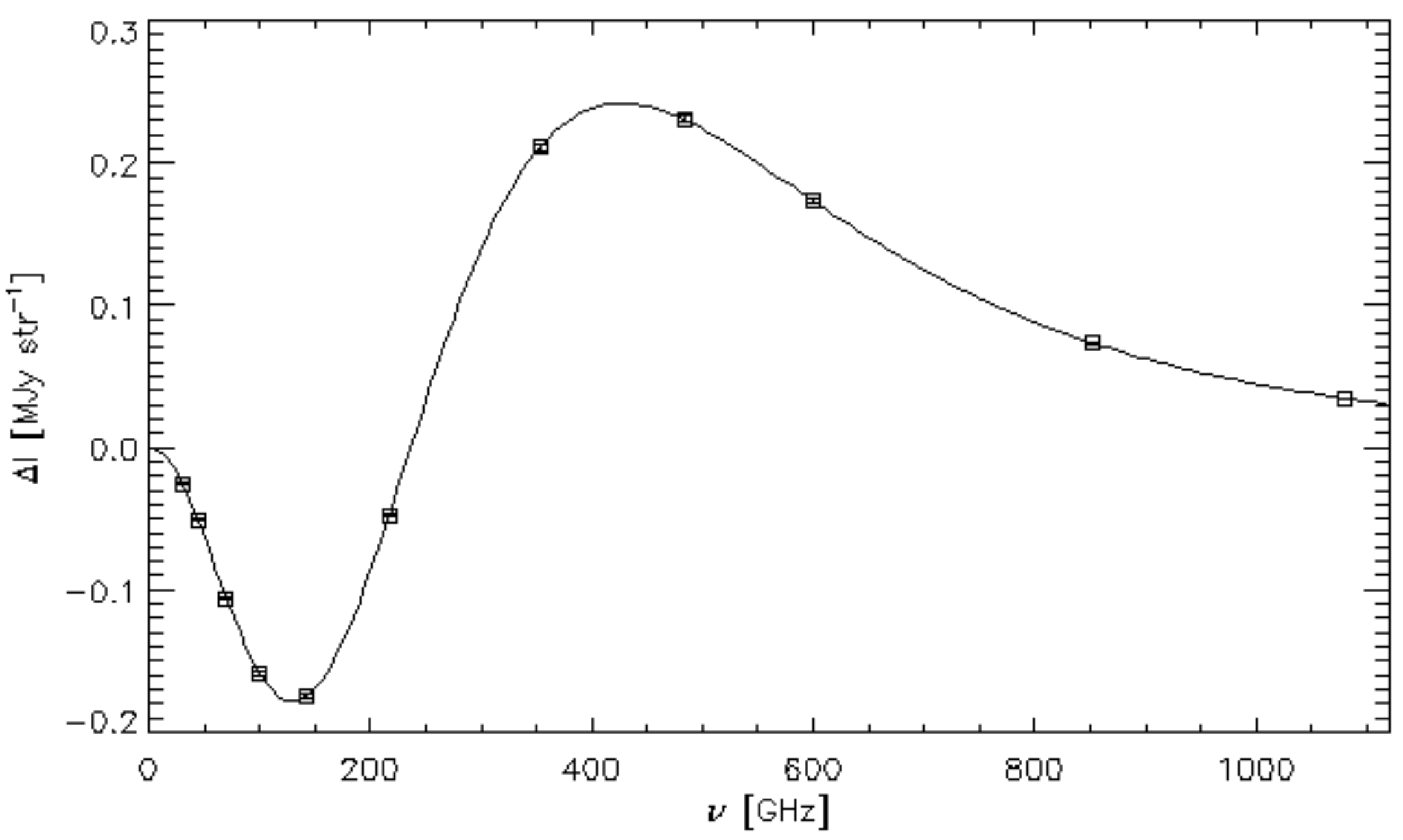}
\caption{
Mock SZ observations along a LOS through a shock extracted 
from a merging cluster \NBODYHYD simulation 
assuming 1\% error in the SZ amplitudes (squares with error bars).
The temperature distribution in this LOS is shown with a solid line 
in Figure~\ref{F:TKEVSHOCK}.
We adopted a \JUTTNER electron velocity distribution to generate 
the SZ signal (solid line; see text for details).
\label{F:SZSHOCK}
}
\end{figure}

We expected that fitting two temperature models to SZ data points at 
only four frequencies would not constrain the EV distribution functions well, 
thus we assumed measurements at the \PLANCK/\HERSCH frequencies
($\nu =$ 30, 44, 70, 100, 143, 217, 353, 545, 600, and 857 GHz).
Carrying out Monte Carlo simulations assuming 1\% error in the SZ amplitudes, 
we found that the distributions of the best-fit temperatures were not well separated, 
thus these set of frequencies do not make it possible to distinguish between
\JUTTNER and modified \JUTTNER EV distributions.
We repeated our simulations adding more measurements around the steep slope
at frequencies between 217 and 353 GHz (244.4, 272.8, 06.9 GHz), but that
did not improve much the accuracy in the best-fit temperatures. 
We found that adding 1080 GHz, to the \PLANCK/\HERSCH frequencies 
to capture the fall off of the SZ signal at very high frequencies
improved on the accuracy in the best-fit temperatures significantly.
The SZ signal from a low temperature gas (a few keV) at very high frequencies 
(above $\sim\;$900 GHz) is negligible, but provide strong constraints on 
high temperature gas.
We show the simulated SZ observations at these extended \PLANCK/\HERSCH 
frequencies and the underlying SZ model through a LOS containing a shock 
in Figure~\ref{F:SZSHOCK} (squares with error bars and solid line).

%
%

We fitted a two temperature model to the mock SZ observations
of the LOS through the shocked region, 
but found that the temperature of the pre-shocked gas 
was not constrained well. 
However, since the temperature of the pre-shocked gas in the LOS
through the shock is very close to that in the LOS through
the pre-shocked gas (see Figure~\ref{F:TKEVSHOCK}),
we fixed the temperature of the pre-shocked gas we 
derived using the LOS through a nearby pre-shocked region.
The best fit value in a LOS through the pre-shocked region                          
was 2.78 keV with a $\pm\;$3\% error,
which is a very good match with the average value derived directly
from our \FLASH simulations (2.68 keV). It is also very close to the
temperature of the pre-shocked gas in the LOS through the shock 
derived from our hydrodynamical simulation (2.65 keV).
Therefore we fixed the lower temperature component at 2.78 keV,
and fit SZ models based on the \JUTTNER and modified \JUTTNER EV 
distributions to the temperature of the shocked gas. 
We show the distributions of the best-fit temperatures
assuming \JUTTNER and modified \JUTTNER EV distributions 
with $\eta = 1$ and $\eta = 2$ in Figure~\ref{F:FITSHOCK} 
(solid, dashed, and dash-dotted histograms).

Again, we obtain good fits for both models with different best-fit electron 
temperatures for the shocked region, 
$T_{J} = 42.1 \pm 0.88$ keV,   
$T_{MJ1} = 45.0 \pm 0.99\;$keV,  
and $T_{MJ2} = 48.2 \pm 1.12\;$keV, 
assuming \JUTTNER and modified \JUTTNER EV distributions with $\eta = 1$ and 2.
The best-fit shock temperature,  $T_{J} = 42.1\;$keV is 
a very good match with the average temperature we obtained 
directly from our \FLASH simulation (42.1$\;$keV).
As we can see from Figure~\ref{F:FITSHOCK}, 
the probability distributions for the best-fit temperatures assuming
\JUTTNER ($P_{J}[T]$) and modified \JUTTNER $P_{MJ1}[T]$, $P_{MJ2}[T]$) 
velocity distributions are clearly separated, suggesting that a 1\% error in the 
SZ measurements would allows us to distinguish between these 
distribution functions.

\section{Discussion}
\label{S:Discussion}

The low density, high temperature ICG in clusters of galaxies 
provides a unique laboratory to test the proposed EV distribution functions.
The temperature in the ICG is not high enough for 
particle pair creation and annihilation to be important,
thus we expect that an equilibrium particle velocity distribution function 
based on particle number conservation can be found.
Also, the assumption of local thermodynamical equilibrium (LTE), 
which is a fundamental criterion for the existence of an 
equilibrium particle velocity distribution, should be valid 
in the ICG in relaxed galaxy clusters. 
The LTE should also be valid in merging
clusters, because the time scale for the EV distribution to reach 
equilibrium is shorter than the time scale for galaxy cluster merging 
(e.g., \citealt{ProkhorovET2011AA529}).

Prokhorov et al. (2011) investigated the possibility to constrain the EV distribution 
in the ICG based on multi-frequency SZ observations. 
They assumed a temperature of 15.3 keV for the ICG, appropriate 
for a high-mass galaxy cluster and considered Maxwellian 
and its relativistic generalization, the \JUTTNER EV distribution functions.
Prokhorov et al. expanded the velocity distribution functions in Fourier series and 
used the Wright formalism for the relativistic (\JUTTNER) distribution  to 
derive equations for the Fourier coefficients. 
Comparing the coefficients, they concluded that the Maxwellian and \JUTTNER 
velocity distributions can be distinguished if the SZ amplitude is measured 
with 0.1\% accuracy. They verified their conclusion with Monte Carlo simulations.
Prokhorov et al. also pointed out the importance of high-frequency SZ observations
(e.g. 375--860 GHz) in constraining EV distribution functions based on the SZ effect.

The Bullet cluster is one of the high-infall velocity merging galaxy clusters, 
which provided the first direct evidence for the existence of dark matter based 
on gravitational lensing \citep{CloweET2006ApJ648}.
The offsets between the mass surface density centers derived from lensing
and the corresponding X-ray emission peaks marking centers of the gas (baryonic) 
components were significant (200--300 kpc, e.g., \citealt{ParaficzET2016AA594}).
Multi-frequency radio/submm observations of the Bullet cluster are available
at four frequencies from 150 GHz to 857 GHz, which makes it an excellent target 
for SZ studies (see Section~\ref{S:SZBULLET}).

%
%
\begin{figure}[t]
\includegraphics[width=.475\textwidth]{\FIGURES/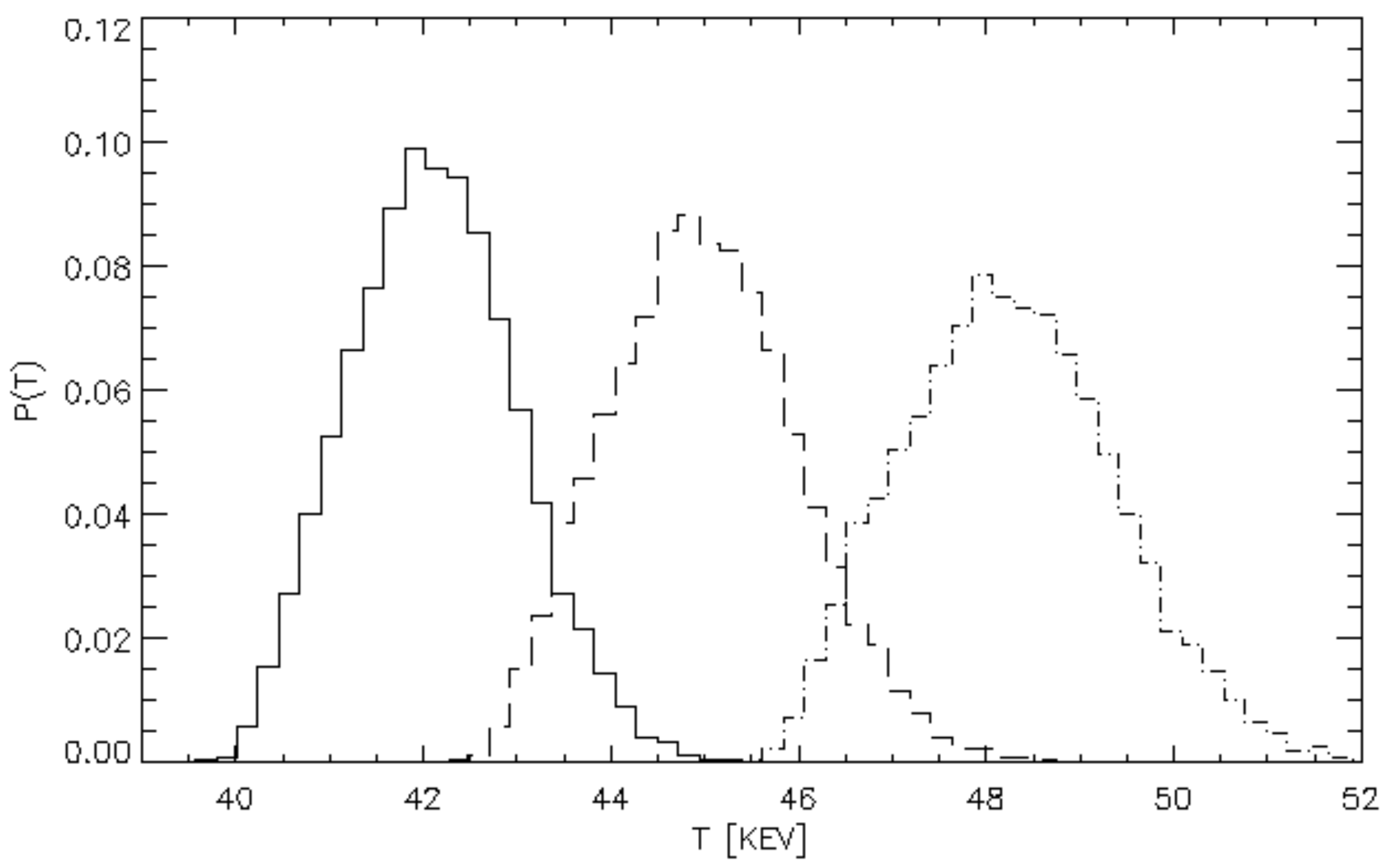}
\caption{
Probability distributions of best-fit shock temperatures from fitting 
SZ models based on the \JUTTNER and modified \JUTTNER
electron velocity distributions with $\eta = 1$ and 2 
(solid, dashed, and dash-dotted lines)
to mock observations of a LOS through a shock extracted from a
merging cluster \NBODYHYD simulation (shown in Figure~\ref{F:SZSHOCK}).
\label{F:FITSHOCK}
}
\end{figure} 

\cite{Cola2011AA527} fitted models to SZ observations of the Bullet cluster 
assuming a single and a double temperature thermal models for the ICG,
and a sum of a single temperature model and one with a non-thermal EV distribution. 
They assumed a \JUTTNER EV distribution for their 
thermal models. Colafrancesco et al. found that a thermal model with a temperature
of $T = 22$ keV and optical depth $\tau = 8.3 \times 10^{-3}$, 
and a model concisting a sum of one thermal and one non-thermal EV models 
provide the best fits (no significant difference in the reduced $\chi^2$). 
They fixed the temperature of the later model at the value derived from X-ray 
measurements ($T_X = 13.9 \pm 0.7$ keV; \citealt{Govoni2004}).
Colafrancesco et al. argued that their later model with a sum of thermal and 
non-thermal plasma is more plausible because:
1) observations suggest the existence of a non-thermal electron component 
in the Bullet cluster \citep{AjelloET2010,PetrosianET2006}, and 
2) their single component thermal model has a best-fit temperature of $T = 22$ keV,
which is much higher than the one derived from X-ray observations 
assuming a single temperature model ($T_X = 13.9$ keV).

\CHANDRA observations of the Bullet cluster found that the projected 
temperature in the Bullet cluster ranges between 4 and 30 keV, 
but the projected temperature in the LOS of the SZ center is about 15 keV
(\citealt{Govoni2004,Markevitch2006ESASP}).
We expect that in the LOS going through the SZ centroid the ICG has a wide 
range of temperatures. In merging clusters, the physical gas temperature can be 
substantially reduced due to projection effects, 
as it was demonstrated quantitatively using hydrodynamical 
simulations (see Figure 4 in \citealt{MolnarBroadhurst2017}).

In order to demonstrate our new method,
we fitted single temperature EV models to SZ observations of the Bullet cluster, 
and obtained best-fit temperatures of $T_{MX} = 18$ keV, $T_{J} = 22$ keV, 
$T_{MJ1} = 23$ keV, and $T_{MJ2} = 24$ keV
assuming Maxwellian, \JUTTNER, modified \JUTTNER EV distributions 
with $\eta = 1$ and 2 (see Section~\ref{S:SZBULLET}).
The best-fit temperature, $T_{J} = 22$ keV, of our single-temperature
model assuming the \JUTTNER EV distribution (Section~\ref{S:SZBULLET})
agrees with that obtained by \cite{Cola2011AA527} 
assuming a single temperature gas adopting the same velocity distribution function.
All three models provided a good fit with no significant difference in 
their $\chi^2$ values ($\Delta \chi^2 < 1$).

Even though the differences in the fitted temperatures between SZ models
based on the Maxwellian EV distribution 
and those using \JUTTNER and modified \JUTTNER 
distributions are as large as $T_{J} - T_{MX} = 4$ keV, 
$T_{MJ1} - T_{MX} = 5$ keV, and $T_{MJ2} - T_{MX} = 6$ keV,
the error bars on these temperatures are even larger ($\sim\,$5 -- 6 keV),
thus we cannot distinguish between these EV models 
based on the best-fit temperatures.

The projected temperature from X-ray observations of the Bullet cluster 
through the SZ centroid is 15 keV, which would be consistent with the
temperature we derived from the SZ observations assuming Maxwellian EV distribution 
(and more than 1$\sigma$ smaller than the temperatures derived assuming its relativistic 
generalizations), but this X-ray temperature is not reliable due to projection effects
(e.g., \citealt{MolnarBroadhurst2017}).
The SZ signal was derived from these observations along a LOS through the center of the 
Bullet cluster, which may be contaminated by a high energy, nonthermal electron
population. Dedicated SZ observations through a LOS which has no 
contamination by nonthermal electrons might simplify the modeling of the cluster
SZ signal. However, the bullet cluster is a large-infall velocity merging cluster, 
thus its LOS temperature structure can only be determined by dedicated
\NBODYHYD simulations. The numerical simulations could provide
a more realistic model for the LOS distribution of the density and temperature 
in the ICG to derive the SZ effect amplitude as a function of frequency.

We obtained similar results fitting the same EV distribution models to mock SZ 
observations adopting a fiducial model based on the \JUTTNER
distribution with electron temperature of 22 keV 
(motivated by our results from fitting models to SZ observations of the Bullet cluster)
assuming a few percent error in the observations (Section~\ref{SS:FITPLASMA}).
The best fit temperatures assuming \JUTTNER, modified \JUTTNER, and
Maxwellian EV distributions were  $T_{J} = 22$ keV, $T_{MJ1} = 23$ keV, 
$T_{MJ2} = 24$ keV, 
and $T_{MX} = 18$ keV, in agreement with our fits to the Bullet cluster data,
but the error bars for the temperatures were smaller.
We noticed, that the best-fit temperatures may differ as much as 1 -- 6 keV 
depending on which model we assume for the velocity distribution.
These results suggest that, if we can derive the ICG temperature from a different,
independent method, we may be able to distinguish between the proposed EV distributions. 
Based on our results, we propose a new method to constrain the electron
velocity distribution in the ICG making use of the frequency 
distribution of the SZ effect. Our method consists of two steps:
1) derive the temperature from fitting models to SZ observations based on different
EV distribution functions; 
2) compare the derived temperatures to a temperature obtained using an independent 
(e.g., X-ray) method.

Such independent method may be provided by 
high-spectral and spectral resolution X-ray observations, which 
could measure accurate temperatures in the ICG based on emission lines.
This method would have less projection bias than the conventional one 
using low spectral-resolution wide spectra. 
In principle, measuring thermal line broadening would allow us to
derive gas temperatures in clusters, but it may be difficult to separate it from broadening 
due to turbulence or resonant line scattering (e.g., \citealt{InogamovSun2003,Molnar2016}).
Line ratios may provide a better diagnostic of gas temperatures
in clusters (e.g., \citealt{NevalainenET2003}). 
Note, however, that models for X-ray emission lines should be 
calculated consistently, using the appropriate EV distribution functions 
(e.g., using the method developed by \citealt{ProkhorovET2009AA496}).

We carried out Monte Carlo simulations of SZ measurements 
of a single-temperature gas to estimate the accuracy needed to 
distinguish between different EV distribution functions.
Fitting to mock SZ observations assuming a 5\% error in the SZ measurements
we found that the relativistic generalizations of the Maxwell velocity distribution,
cannot be distinguished from each other, 
but they can be distinguished from the Maxwellian velocity distribution, 
since the probability distribution of best-fitted temperatures assuming 
Maxwellian velocity distribution, $P_{MX}(T)$, is clearly separated from those
based on the \JUTTNER and modified \JUTTNER distributions with $\eta = 1$ and 2, 
$P_{J}(T)$, $P_{MJ1}(T)$, and $P_{MJ2}(T)$ (see Figure~\ref{F:MC5PER}).
Assuming that the temperature can be derived using another method 
with the same accuracy as the one based on the SZ observations 
assuming 5\% errors in the SZ amplitudes ($\sim 1.25$ keV), 
we can still distinguish between the Maxwellian and its relativistic generalizations, 
the \JUTTNER, and the modified \JUTTNER distributions with high significance ($\sim 2\sigma$).
In order to distinguish between different relativistic generalizations of the 
Maxwell velocity distribution we need more accurate SZ measurements than 5\%.
The significance of measuring SZ effect with 5\% accuracy is that 
it could justify the usage of relativistic generalizations 
of the Maxwell velocity distributions in studying the ICG in galaxy clusters.

%
%
\begin{figure}[t]
\includegraphics[width=.47\textwidth]{\FIGURES/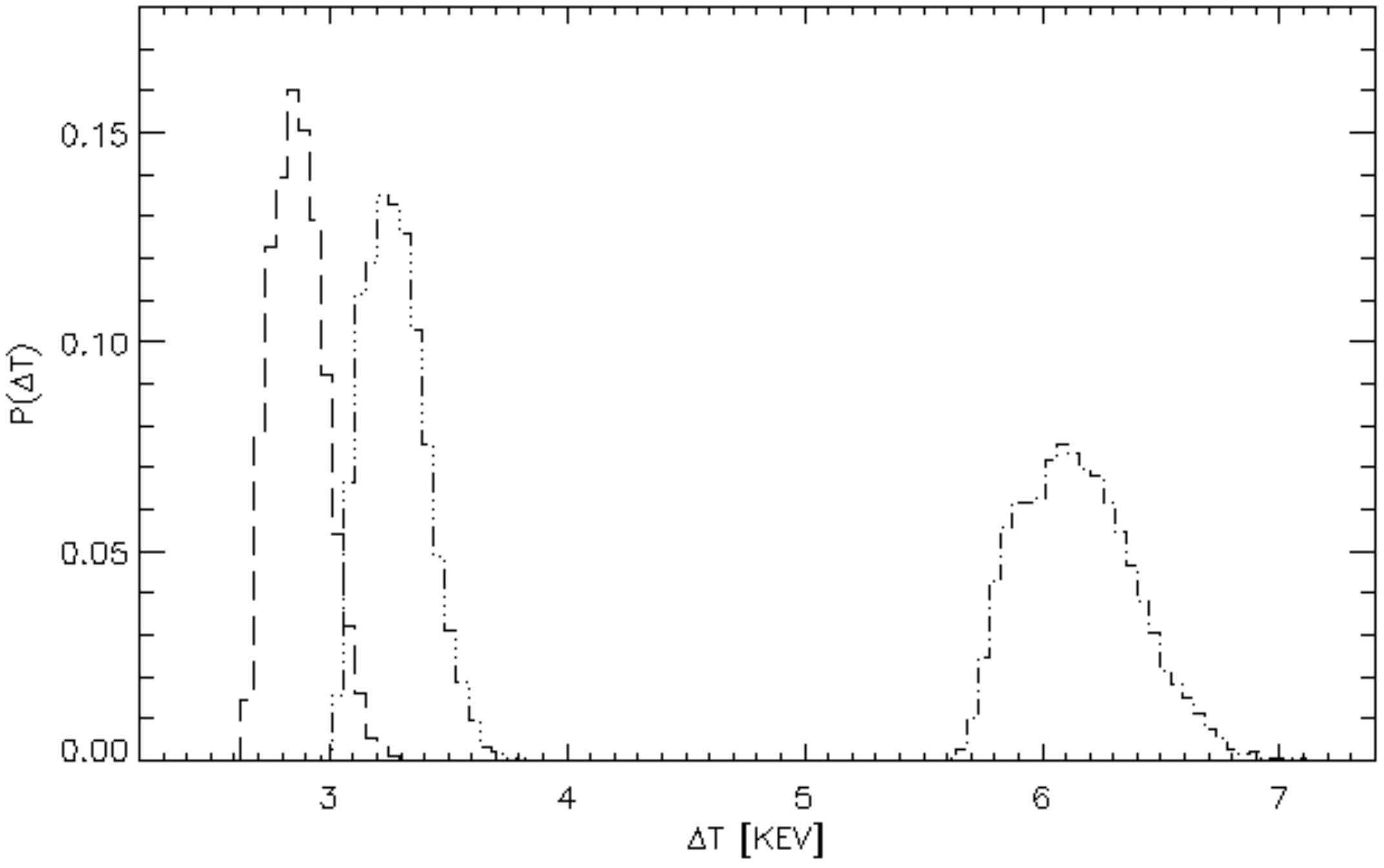}
\caption{
Probability distributions of the differences of the best-fit shock temperatures from fitting 
SZ models to mock observations of a line of sight through a shock extracted from a
merging cluster simulation assuming 1\% errors in measurements.
Dashed and dash-dotted lines show differences between best-fit temperatures
based on the \JUTTNER and modified \JUTTNER electron velocity distributions
with $\eta = 1$ ($T_{MJ1} - T_{J}$) and $\eta = 2$ ($T_{MJ2} - T_{J}$).
The dash-dot-dot-dotted line represents temperature differences between 
assuming modified \JUTTNER distributions with $\eta = 1$ and $\eta = 2$ 
($T_{MJ2} - T_{MJ1}$).
\label{F:PDELTSHOCK}
}
\end{figure} 

As we discussed it in Section~\ref{S:RELKIN}, there are difficulties
in developing a self-consistent extension of our nonrelativistic 
kinetic theory to the relativistic regime.
Therefore, it would be of fundamental importance to use the SZ effect to verify 
which relativistic generalization of the Maxwell distribution is correct: 
the \JUTTNER or one of the modified \JUTTNER distributions.
This would provide us an empirical test for 
different relativistic generalizations of our non-relativistic kinetic theory.
Fitting to mock SZ observations as before, but assuming 1\% errors in the 
SZ measurements we find that the probability distributions of best-fit temperatures 
from fitting SZ models based on the \JUTTNER, $P_{J}(T)$, and 
modified \JUTTNER EV distributions with $\eta = 1$ and $\eta = 2$,
$P_{MJ1}(T)$, and $P_{MJ2}(T)$, to the same simulated 
SZ observations are well separated (Figure~\ref{F:MC1PER}), 
thus we can distinguish them.
Assuming that the measurement error in the 
temperature of the ICG in a cluster is $\pm 0.9\,$keV ($3 \sigma$)
based on another method, our results suggest that we can distinguish between 
the \JUTTNER and modified \JUTTNER EV distributions 
with $\eta = 1$ with high significance ($\sim 3 \sigma$).
Measuring the gas temperature with an error of $\pm 1.8\,$keV ($3 \sigma$)
with another method, we expect that we can distinguish between 
the \JUTTNER and modified \JUTTNER distribution with $\eta = 2$ 
with high significance ($\sim 3 \sigma$).
In order to distinguish between modified \JUTTNER distributions with
$\eta = 1$ and $\eta = 2$, we would need an accuracy of about $\pm 1.0\,$keV.
Note, that 1\% errors in the SZ measurements are already achievable 
with some instruments (e.g., \HERSCH-SPIRE; \citealt{GriffinET2010AA518}).

We demonstrated that our method can be used to distinguish between different
EV distribution functions if the gas has a single temperature.
However, any LOS through a galaxy cluster has a range of temperatures,
even if the cluster is relaxed. 
In order to test our proposed method to EV distribution functions in a more
realistic cluster model, we fitted to mock SZ observations as before 
to a LOS through a shock extracted from our self-consistent \NBODYHYD
simulation assuming 1\% errors in the SZ measurements
at the \PLANCK/ \HERSCH frequencies and an additional high frequency (1080 GHz).
Again, we found that the probability distributions of best-fit temperatures 
from fitting SZ models based on the \JUTTNER, $P_{J}(T)$, and 
modified \JUTTNER EV distributions with $\eta = 1$ and 2, 
$P_{MJ1}(T)$ and $P_{MJ2}(T)$, to simulated SZ observations are 
well separated (Figure~\ref{F:FITSHOCK}), thus we can distinguish them.
We show the probability distribution of the differences in the best-fit 
temperatures in a LOS through a shock using EV distributions of the form of  
\JUTTNER and modified \JUTTNER with $\eta = 1$ and 2
in Figure~\ref{F:PDELTSHOCK}.
In this Figure, dashed and dash-dotted lines show the
distribution of best-fit temperature differences between assuming 
\JUTTNER and modified \JUTTNER EV distributions with $\eta = 1$ ($T_{MJ1} - T_{J}$) 
and $\eta = 2$ ($T_{MJ2} - T_{J}$).
Best-fit temperature differences between based on modified \JUTTNER distributions 
with $\eta = 1$ and $\eta = 2$ ($T_{MJ2} - T_{MJ1}$) are displayed with a
dash-dot-dot-dotted line.
We find that assuming that the $3\sigma$ measurement error in the temperature 
of the ICG in a cluster is $\pm 2.65\,$keV ($\pm 5.65\,$keV) based on another method, 
our results suggest that we can distinguish between the \JUTTNER and the modified 
\JUTTNER EV distributions with $\eta = 1$ ($\eta = 2$) with high significance ($\sim 3\sigma$).
In order to distinguish between modified \JUTTNER distributions with $\eta = 1$ 
and $\eta = 2$ with high significance ($\sim 3\sigma$), we would need to measure 
the ICG temperature with an error of $\pm 2.98\,$keV ($3\sigma$) using another method,
similar to the requirement for distinguishing between \JUTTNER and modified 
\JUTTNER distribution with $\eta = 1$.

In general, higher temperature ICG would make it easier the test EV distribution functions. 
We found that, for gas temperatures of $\sim 20\,$keV and $\sim 40\,$keV, the differences 
between the best-fit temperatures assuming \JUTTNER and a modified \JUTTNER EV 
distribution with $\eta = 1$, $T_{MJ1}$ - $T_{J}$, are 4\% and 7\%, while for temperatures 
adopting a modified \JUTTNER distribution with $\eta = 2$, the differences, $T_{MJ2}$ - $T_{J}$, 
are 8\% and 14\%. Based on our results, we expect that for shock temperatures of 
$\sim\,$60 keV, the differences between best-fit temperatures for 
$T_{MJ1}$ - $T_{J}$, $T_{MJ2}$ - $T_{J}$, and  $T_{MJ2}$ - $T_{MJ1}$
would be about 14\%, 21\%, and 15\%, much easier the achieve.

The observational frequencies could be chosen to minimize the errors 
in the temperature derivations based on different EV distribution models.
High-spectral and spatial resolution radio/submm observations 
may help to separate some of the contaminating components
by choosing LOSs avoiding substructures and radio halos and relics
containing high energy nonthermal electrons. 
Radio relics associated with shocks are patchy in most merging clusters, 
thus they can be avoided (e.g., except the ``Sausage cluster'', 
although it can be modeled with \NBODYHYD simulations; 
\citealt{MolnarBroadhurst2017}).
Also, it will be necessary to identify other methods to derive accurate 
temperatures in the ICG.
X-ray emission lines from the ICG may provide the required accuracy.
A dedicated feasibility study would be important  
based on more realistic ICG cluster models 
taking into account the response of the available detectors 
(spectral and spatial resolution, sensitivity, etc.), and 
contaminating effects in conjunction with an analysis of 
other, independent, methods to derive the temperature in the ICG.
We leave this detailed analysis for the future.

\section{Conclusion}
\label{S:Conclusion}

We developed a new method to test relativistic kinetic theories based on
observations of the thermal SZ effect in galaxy clusters.
We demonstrated that the frequency dependence of the SZ effect can be used
to distinguish between different EV equilibrium distribution functions
in the ICG assuming that an independent measurement of the temperature is 
available from another method.
This new method is based on our observation that different EV distribution 
functions result in different temperatures when their models are fitted to the 
same SZ data.

We found that a 5\% accuracy is necessary in the SZ amplitude measurements 
of high single temperature gas ($\sim 20$ keV)
to distinguish between non-relativistic (Maxwellian), and its relativistic 
generalizations, the \JUTTNER and the modified \JUTTNER EV distributions.
In order to identify the correct relativistic generalization of the Maxwell velocity distribution 
as the \JUTTNER or one of the modified \JUTTNER EV distributions, 
we would need about 1\% errors in the measured SZ amplitudes.

We demonstrated that our method works in a simulated merging cluster
using a LOS through a shocked region. 
A LOS through a shock contains two phases of the gas:
shocked and pre-shocked gas, with a range of temperatures. 
We found that the change of the temperature within each phase is relatively small.
We carried out Monte Carlo simulations assuming a 1\% error in the SZ measurements
at the \PLANCK/\HERSCH frequencies and at 1080 GHz.
We found that two temperature gas models based on \JUTTNER and
modified \JUTTNER EV distributions with $\eta = 1$ and 2
fit well to the mock SZ observations.
Our results suggest that these three distributions can be distinguished with high significance
based on their best-fit temperatures to the high temperature $\,\simgreat\, 40$ keV)
shocked gas in clusters of galaxies.
We found that in order to reach high significance in distinguishing 
between \JUTTNER and modified \JUTTNER EV distributions we need 
observations at THz frequencies.
Astrophysics in the THz frequency range is a promising new field covering
a wide range of research topics from black holes to exoplanets and cosmology
\citep[e.g.,][]{GurvitsET2019arXiv1908,Withington2004}.
There are atmospheric windows at e.g., $\sim 1.1\,$THz, and  $\sim 1.5\,$THz,
which make ground-based observations available 
at high altitude, dry observing sites (e.g., from Antarctica \citep{SetaET2013}.
Using heterodyne arrays on board of SOFIA balloon experiment,
observations have already been made between 1.8 THz and 4.7 THz
\citep{RisacherET2018}, and more ground and space based THz telescopes 
are under development \citep[e.g.,][]{GurvitsET2019arXiv1908}.

Our results suggest that it would be worth while to carry out 
a feasibility study of our proposed new method to constrain EV 
distributions in the ICG based on more realistic cluster models derived 
from hydrodynamical cosmological simulations. 
Systematic effects from inhomogeneous temperature distribution,
substructures and nonthermal electron populations could be studied 
using clusters from cosmological simulations.

\acknowledgements
We thank the referee for detailed comments and suggestions,
which helped to improve our paper.
This work was supported in part by the Ministry of Science and
Technology of Taiwan (grants MOST 106-2628-M-001-003-MY3 
and MOST 109-2112-M-001-018-MY3) and by Academia Sinica 
(grant ASIA-107-M01).

%
%
\bibliographystyle{apj}

\begin{thebibliography}{99}


\bibitem[Ajello et al.(2010)]{AjelloET2010}
 Ajello, M., Rebusco, P., Cappelluti, N., et al.\ 2010, \apj, 725, 1688


\bibitem[Aragon-Munoz \& Chacon-Acosta(2018)]{Aragon2018}
Aragon-Munoz, L. \& G. Chacon-Acosta, G.\ 2018, J. Phys.: Conf. Ser. 1030 012004


\bibitem[Baldi et al.(2019)]{BaldiET2019AA630}
 Baldi, A.~S., Bourdin, H., Mazzotta, P., et al.\ 2019, \aap, 630, A121


\bibitem[Bernstein(2004)]{Bernstein2004}
 Bernstein, J.\ 2004, {\it Kinetic Theory in the Expanding Universe}, Cambridge University Press, Cambridge


\bibitem[Birkinshaw(1999)]{Birkinshaw1999}
 Birkinshaw, M.\ 1999, \physrep, 310, 97


\bibitem[Bourdin et al.(2017)]{BourdinET2017ApJ843}
 Bourdin, H., Mazzotta, P., Kozmanyan, A., et al.\ 2017, \apj, 843, 72


\bibitem[Carlstrom et al.(2002)]{CarlstromET2002}
 Carlstrom, J. E, Holder, G. P., \& Reese, E. D., 2002, \araa, 40, 643


\bibitem[Chandrasekhar(1960)]{Chandra1960}
 Chandrasekhar, S.\ 1960, New York: Dover, 1960


\bibitem[Clowe et al.(2006)]{CloweET2006ApJ648}
 Clowe, D., Brada{\v{c}}, M., Gonzalez, A.~H., et al.\ 2006, \apjl, 648, L109


\bibitem[Colafrancesco \& Marchegiani(2010)]{ColaMarc2010AA520}
 Colafrancesco, S., \& Marchegiani, P.\ 2010, \aap, 520, A31


\bibitem[Colafrancesco et al.(2011)]{Cola2011AA527}
 Colafrancesco, S., Marchegiani, P., \& Buonanno, R.\ 2011, \aap, 527, L1


\bibitem[Colafrancesco et al.(2009)]{ColaET2009AA494}
 Colafrancesco, S., Prokhorov, D., \& Dogiel, V.\ 2009, \aap, 494, 1


\bibitem[Cubero et al.(2007)]{CuberoET2007}
 Cubero, D., Casado-Pascual, J., Dunkel, J., et al.\ 2007, \prl, 99, 170601


\bibitem[Dunkel \& H{\"a}nggi(2007)]{DunkHang2007PhyA374}
 Dunkel, J., \& H{\"a}nggi, P.\ 2007, Physica A Statistical Mechanics and its Applications, 374, 559


\bibitem[Dunkel et al.(2009)]{Dunkel2009NatPh}
 Dunkel, J., H{\"a}nggi, P., \& Hilbert, S.\ 2009, Nature Physics, 5, 741


\bibitem[Dunkel et al.(2007)]{DunkelET2007NJPh9}
 Dunkel, J., Talkner, P., \& H{\"a}nggi, P.\ 2007, New Journal of Physics, 9, 144


\bibitem[Fryxell et al.(2000)]{Fryxell2000ApJS131p273}
 Fryxell, B., et al.\ 2000, \apjs, 131, 273 


\bibitem[Gomez et al.(2004)]{Gomez2004}
 Gomez, P., Romer, A.~K., Peterson, J.~B., et al.\ 2004, Plasmas in the Laboratory and in the Universe: New Insights and New Challenges, 361


\bibitem[Govoni et al.(2004)]{Govoni2004}
 Govoni, F., Markevitch, M., Vikhlinin, A., et al.\ 2004, \apj, 605, 695


\bibitem[Gurvits et al.(2019)]{GurvitsET2019arXiv1908}
 Gurvits, L.~I., Paragi, Z., Casasola, V., et al.\ 2019, arXiv e-prints, arXiv:1908.10767


\bibitem[Hakim(2011)]{Hakim2011}
R, Hakim, 2011, ``Introduction to relativistic statistical mechanics : classical and quantum'', World scientific, Singapore


\bibitem[Halverson et al.(2009)]{HalversonET2009}
 Halverson, N.~W., Lanting, T., Ade, P.~A.~R., et al.\ 2009, \apj, 701, 42


\bibitem[Hansen et al.(2002)]{HansenET2002ApJ573}
 Hansen, S.~H., Pastor, S., \& Semikoz, D.~V.\ 2002, \apjl, 573, L69


\bibitem[Horwitz et al.(1973)]{HorwitzPiron1973}
 Horwitz, L.~P., \& Piron, C.\ 1973, Helv. Phys. Acta, 46, 316 


\bibitem[Horwitz et al.(1989)]{HorwitzET1989PhyA161}
 Horwitz, L.~P., Shashoua, S., \& Schieve, W.~C.\ 1989, Physica A Statistical Mechanics and its Applications, 161, 300


\bibitem[Horwitz et al.(1981)]{HorwitzET1981}
 Horwitz, L.~P., Schieve, W.~C., \& Piron, C.\ 1981, Annals of Physics, 137, 306


\bibitem[Griffin et al.(2010)]{GriffinET2010AA518}
 Griffin, M.~J., Abergel, A., Abreu, A., et al.\ 2010, \aap, 518, L3


\bibitem[Inogamov \& Sunyaev(2003)]{InogamovSun2003}
 Inogamov, N.~A., \& Sunyaev, R.~A.\ 2003, Astronomy Letters, 29, 791


\bibitem[J{\"u}ttner(1911)]{Juttner1911}
 J{\"u}ttner, F.\ 1911, Annalen der Physik, 339, 856


\bibitem[Kaniadakis(2006)]{Kaniadakis2006}
 Kaniadakis, G.\ 2006, Physica A Statistical Mechanics and its Applications, 365, 17
[197] G. Kaniadakis. Towards a relativistic statistical theory. Physica A, 365:17–23, 2006


\bibitem[Kompaneets(1957)]{Komp1957}
 Kompaneets, A.~S.\ 1957, Soviet Journal of Experimental and Theoretical Physics, 4, 730


\bibitem[Lehmann(2006)]{Lehmann2006JMP47}
 Lehmann, E.\ 2006, Journal of Mathematical Physics, 47, 023303


\bibitem[Leutwyler(1965)]{Leutwyler1965}
 Leutwyler, H.\ 1965, Il Nuovo Cimento, 37, 556


\bibitem[Markevitch(2006)]{Markevitch2006ESASP}
 Markevitch, M.\ 2006, The X-ray Universe 2005, 723


\bibitem[Marmo et al.(1984)]{MarmoET1984}
 Marmo, G., Mukunda, N., \& Sudarshan, E.~C.~G.\ 1984, \prd, 30, 2110


\bibitem[Molnar et al.(2012)]{MolnarET2012ApJ748}
 Molnar, S.~M., Hearn, N.~C., \& Stadel, J.~G.\ 2012, \apj, 748, 45 


\bibitem[Molnar(2015)]{Molnar2015}
 Molnar, S.~M., 2015, {\it Cosmology with Clusters of Galaxies}, 
 Nova Science Publishers, New York


\bibitem[Molnar(2016)]{Molnar2016}
 Molnar, S.\ 2016, Front. Astron. Space Sci., 2, 7


\bibitem[Molnar \& Birkinshaw(1999)]{MolnarBirk1999}
 Molnar, S.~M., \& Birkinshaw, M.\ 1999, \apj, 523,78


\bibitem[Molnar \& Broadhurst(2015)]{MolnarBroadhurst2015}
 Molnar, S.~M., \& Broadhurst, T.\ 2015, \apj, 800, 37 
 
 
\bibitem[Molnar \& Broadhurst(2017)]{MolnarBroadhurst2017}
 Molnar, S.~M., \& Broadhurst, T.\ 2017, \apj, 841, 46 


\bibitem[Molnar \& Broadhurst(2018)]{MolnarBroadhurst2018}
 Molnar, S.~M., \& Broadhurst, T.\ 2018, \apj, 862, 112 


\bibitem[Montakhab et al.(2009)]{MontakhabET2009}
 Montakhab, A., Ghodrat, M., \& Barati, M.\ 2009, \pre, 79, 031124


\bibitem[Nevalainen et al.(2003)]{NevalainenET2003}
 Nevalainen, J., Lieu, R., Bonamente, M., et al.\ 2003, \apj, 584, 716


\bibitem[Nozawa et al.(1998)]{NozawaET1998ApJ507}
 Nozawa, S., Itoh, N., \& Kohyama, Y.\ 1998, \apj, 507, 530


\bibitem[Paraficz et al.(2016)]{ParaficzET2016AA594}
 Paraficz, D., Kneib, J.-P., Richard, J., et al.\ 2016, \aap, 594, A121


\bibitem[Peano et al.(2009)]{PeanoET2009}
 Peano, F., Marti, M., Silva, L.~O., et al.\ 2009, \pre, 79, 025701


\bibitem[Petrosian et al.(2006)]{PetrosianET2006}
 Petrosian, V., Madejski, G., \& Luli, K.\ 2006, \apj, 652, 948


\bibitem[Plagge et al.(2010)]{PlaggeET2010}
 Plagge, T., Benson, B.~A., Ade, P.~A.~R., et al.\ 2010, \apj, 716, 1118


\bibitem[Planck Collaboration et al.(2016)]{PLANCKXXII2016}
 Planck Collaboration, Aghanim, N., Arnaud, M., et al.\ 2016, \aap, 594, A22


\bibitem[Pointecouteau et al.(1998)]{PointET1998AA336}
 Pointecouteau, E., Giard, M., \& Barret, D.\ 1998, \aap, 336, 44


\bibitem[Prokhorov et al.(2011)]{ProkhorovET2011AA529}
 Prokhorov, D.~A., Colafrancesco, S., Akahori, T., et al.\ 2011, \aap, 529, A39


\bibitem[Prokhorov et al.(2010)]{ProkET2010AA524}
 Prokhorov, D.~A., Dubois, Y., \& Nagataki, S.\ 2010, \aap, 524, A89


\bibitem[Prokhorov et al.(2009)]{ProkhorovET2009AA496}
 Prokhorov, D.~A., Durret, F., Dogiel, V., et al.\ 2009, \aap, 496, 25


\bibitem[Rephaeli(1995)]{Rephaeli1995ApJ445}
 Rephaeli, Y.\ 1995, \apj, 445, 33


\bibitem[Ricker(2008)]{Ricker2008ApJS176}
 Ricker, P.~M.\ 2008, \apjs, 176, 293 


\bibitem[Risacher et al.(2018)]{RisacherET2018}
 Risacher, C., G{\"u}sten, R., Stutzki, J., et al.\ 2018, Journal of Astronomical Instrumentation, 7, 1840014


\bibitem[Schieve(2005)]{Schieve2005}
 Schieve, W.~C.\ 2005, Foundations of Physics, 35, 1359


\bibitem[Seta et al.(2013)]{SetaET2013}
Seta, M., Nakai, N., Shun Ishii, S., et al.\ 2013, in Astrophysics from Antarctica, Proceedings IAU Symposium No. 288, 2012, eds.: Burton, M., G., Cui, X,  \& Tothill, N. F. H.


\bibitem[Sunyaev \& Zeldovich(1980)]{SZ1980ARAA18}
 Sunyaev, R.~A., \& Zeldovich, I.~B.\ 1980, \araa, 18, 537


\bibitem[Synge(1957)]{Synge1957}
Synge, J. L., 1957, {\it The Relativistic Gas}, North-Holland, Amsterdam


\bibitem[van Hees et al.(2006)]{HeesET2006PhRvC}
 van Hees, H., Greco, V., \& Rapp, R.\ 2006, \prc, 73, 034913


\bibitem[Withington(2004)]{Withington2004}
 Withington, S.\ 2004, Philosophical Transactions of the Royal Society of London Series A, 362, 395


\bibitem[Wright(1979)]{Wright1979}
 Wright, E.~L.\ 1979, \apj, 232, 348 


\bibitem[Zemcov et al.(2010)]{ZemcovET2010AA518}
 Zemcov, M., Rex, M., Rawle, T.~D., et al.\ 2010, \aap, 518, L16



\end{thebibliography}


\end{document}